\documentclass{emulateapj}

\newcommand{\boldsymbol}[1]{\mbox{\boldmath{${#1}$}}}

\usepackage{natbib}
\usepackage[usenames]{color}

\newcommand{\bd}{\begin{displaymath}}
\newcommand{\ed}{\end{displaymath}}
\newcommand{\be}{\begin{equation}}
\newcommand{\ee}{\end{equation}}
\newcommand{\beaa}{\begin{eqnarray*}}
\newcommand{\eeaa}{\end{eqnarray*}}
\newcommand{\bea}{\begin{eqnarray}}
\newcommand{\eea}{\end{eqnarray}}

\newcommand\rxj{RXJ1131$-$1231}
\newcommand\blens{B1608$+$656}

\def\hst{\textit{HST}}

\def\zd{z_{\rm d}}
\def\zs{z_{\rm s}}

\def\Ok{\Omega_{\rm k}}
\def\Ode{\Omega_{\rm de}}
\def\Om{\Omega_{\rm m}}
\def\OL{\Omega_{\Lambda}}

\def\parAllVec{\boldsymbol{{\xi}}}
\def\parCosVec{\boldsymbol{{\pi}}}
\def\parLensVec{\boldsymbol{{\eta}}}

\def\slope{\gamma'}
\def\thEin{\theta_{\rm E}}
\def\kext{\kappa_{\rm ext}}
\def\gext{\gamma_{\rm ext}}
\def\tdist{D_{\Delta t}}
\def\tdistmod{D_{\Delta t}^{\rm model}}
\def\rani{r_{\rm ani}}

\def\thEinS{\theta_{\rm E,S}}

\def\dt{\Delta t}
\def\dtVec{\boldsymbol{\Delta t}}

\def\dacsVec{\boldsymbol{d_{\rm ACS}}}
\def\dacsVecP{\boldsymbol{d_{\rm ACS}^{\rm P}}}
\def\dacsi{d_{{\rm ACS,}i}}
\def\dacsiP{d_{{\rm ACS,}i}^{\rm P}}

\def\vdisp{\sigma}

\def\denvVec{\boldsymbol{d_{\rm env}}}

\def\imVec{\boldsymbol{\theta}}
\def\im{\theta}
\def\srVec{\boldsymbol{\beta}}
\def\sr{\beta}

\def\esrVec{\boldsymbol{s}}

\def\agnVec{\boldsymbol{\nu}}

\def\eagnVec{\boldsymbol{a}}

\def\galVec{\boldsymbol{g}}

\def\parsVec{\boldsymbol{\eta}}

\def\Bmat{\boldsymbol{\mathsf{B}}}
\def\Lmat{\boldsymbol{\mathsf{L}}}

\def\kms {\rm km\,s^{-1}}
\def\kmsMpc {\rm km\,s^{-1}\,Mpc^{-1}}

\def\Dtfitlam{1425.6}
\def\Dtfitmu{6.4993}
\def\Dtfitsig{0.19377}

\def\hypG{\mathsf{H}^{\rm global}}
\def\hypI{\mathsf{H}^{\rm ind}}
\def\dR{\boldsymbol{d}^{\rm R}}
\def\dB{\boldsymbol{d}^{\rm B}}

\newcommand{\sref}[1]{Section~\ref{#1}}
\newcommand{\fref}[1]{Figure~\ref{#1}}
\newcommand{\tref}[1]{Table~\ref{#1}}
\newcommand{\eref}[1]{Equation~(\ref{#1})}

\newcommand{\degt}{\mathrm{deg}}

\def\eg{{e.g.,}}
\def\ie{{i.e.,}}

\begin{document}

\title{Two accurate time-delay distances from strong lensing: implications for cosmology}

\author{S.~H.~Suyu\altaffilmark{1,2,3},
M.~W.~Auger\altaffilmark{4},
S.~Hilbert\altaffilmark{2,5},
P.~J.~Marshall\altaffilmark{6},
M.~Tewes\altaffilmark{7},
T.~Treu\altaffilmark{1,*},
C.~D.~Fassnacht\altaffilmark{8},
L.~V.~E.~Koopmans\altaffilmark{9},
D.~Sluse\altaffilmark{10}, 
R.~D.~Blandford\altaffilmark{2},
F.~Courbin\altaffilmark{7},
G.~Meylan\altaffilmark{7}}
\altaffiltext{1}{Department of Physics, University of California, Santa Barbara, CA 93106, USA} 
\altaffiltext{2}{Kavli Institute for Particle Astrophysics and Cosmology, Stanford University, 452 Lomita Mall, Stanford, CA 94035, USA} 
\altaffiltext{3}{Institute of Astronomy and Astrophysics, Academia Sinica, P.O.~Box 23-141, Taipei 10617, Taiwan}
\altaffiltext{4}{Institute of Astronomy, University of Cambridge, Madingley Rd, Cambridge, CB3 0HA, UK}
\altaffiltext{5}{Max-Planck-Institut f{\"u}r Astrophysik, Karl-Schwarzschild-Str.~1, 85748 Garching, Germany}
\altaffiltext{6}{Department of Physics, University of Oxford, Keble Road, Oxford, OX1 3RH, UK}
\altaffiltext{7}{Laboratoire d'Astrophysique, Ecole Polytechnique F{\'e}d{\'e}rale de Lausanne (EPFL), Observatoire de Sauverny, CH-1290 Versoix, Switzerland}
\altaffiltext{8}{Department of Physics, University of California, Davis, CA 95616, USA}
\altaffiltext{9}{Kapteyn Astronomical Institute, University of Groningen, P.O.Box 800, 9700 AV Groningen, The Netherlands}
\altaffiltext{10}{Argelander-Institut f\"ur Astronomie, Auf dem H\"ugel 71, 53121 Bonn, Germany}
\altaffiltext{*}{Packard Research Fellow}

\email{suyu@asiaa.sinica.edu.tw}

\shorttitle{Cosmological constraints from time-delay lenses}
\shortauthors{Suyu et al.}


\begin{abstract}

Strong gravitational lenses with measured time delays between the
multiple images and models of the lens mass distribution allow a
one-step determination of the time-delay distance, and thus a
measure of cosmological parameters.  We present a blind analysis of
the gravitational lens \rxj\ incorporating (1) the newly measured time
delays from COSMOGRAIL, the COSmological MOnitoring of GRAvItational
Lenses, (2) archival \textit{Hubble Space Telescope} imaging of the
lens system, (3) a new velocity-dispersion measurement of the lens
galaxy of $323\pm20\,\kms$ based on Keck spectroscopy, and (4) a
characterization of the line-of-sight structures via observations of
the lens' environment and ray tracing through the Millennium
Simulation.  Our blind analysis is designed to prevent experimenter
bias. The joint analysis of the data sets allows a time-delay distance
measurement to 6\% precision that takes into account all known
systematic uncertainties.  In combination with the
\textit{Wilkinson Microwave Anisotropy Probe} seven-year (WMAP7) data
set in flat $w$CDM cosmology, our unblinded cosmological constraints
for \rxj\ are: $H_0=80.0^{+5.8}_{-5.7}\,\kmsMpc$, $\Ode=0.79\pm0.03$,
$w=-1.25^{+0.17}_{-0.21}$.  We find the results to be statistically
consistent with those from the analysis of the gravitational lens
\blens, permitting us to combine the inferences from these two lenses.  
The joint constraints from the two lenses and WMAP7 are
$H_0=75.2^{+4.4}_{-4.2}\,\kmsMpc$, $\Ode=0.76^{+0.02}_{-0.03}$ and $w
= -1.14^{+0.17}_{-0.20}$ in flat $w$CDM, and
$H_0=73.1^{+2.4}_{-3.6}\,\kmsMpc$, $\OL=0.75^{+0.01}_{-0.02}$ and
$\Ok=0.003^{+0.005}_{-0.006}$ in open $\Lambda$CDM.  Time-delay lenses
constrain especially tightly the Hubble constant $H_0$ (5.7\% and 4.0\%
respectively in $w$CDM and open $\Lambda$CDM) and curvature of the
universe. The overall information content is similar to that of
Baryon Acoustic Oscillation experiments. Thus, they complement well
other cosmological probes, and provide an independent check of unknown
systematics.  Our measurement of the Hubble constant is 
completely independent of those based on the local
distance ladder method, providing an important consistency check of
the standard cosmological model and of general relativity.

\end{abstract}



\section{Introduction} 
\label{sec:intro}

In the past century precise astrophysical measurements of the geometry
and content of the universe (hereafter cosmography) have led to some
of the most remarkable discoveries in all of physics. These include
the expansion and acceleration of the Universe, its large scale
structure, and the existence of non-baryonic dark matter
\citep[see review by][]{FreedmanTurner03}. These observations form the
empirical foundations of the standard cosmological model, which is
based on general relativity and the standard model of particle physics
but requires additional non-standard features such as 
non-baryonic dark matter and dark energy.

Even in the present era of so-called precision cosmography, many
profound questions about the Universe remain unanswered. What is the
nature of dark energy? What are the properties of the dark matter
particle? How many families of relativistic particles are there? What
are the masses of the neutrinos? Is general relativity the correct
theory of gravity? Did the Universe undergo an inflationary phase in
its early stages?

From an empirical point of view, the way to address these questions is
to increase the accuracy and precision of cosmographic
experiments. For example, clues about the nature of dark energy can be
gathered by measuring the expansion history of the Universe to very
high precision, and modeling the expansion as being due to a dark
energy component having an equation of state parametrized by $w$ that
evolves with cosmic time \citep[\eg][and references therein]{FTH08}.
Likewise, competing inflationary models can be tested by measuring the
curvature of the Universe to very high precision.  Given the high
stakes involved, it is essential to develop multiple independent
methods as a way to control for known systematic uncertainties,
uncover new ones,
and ultimately discover
discrepancies that may reveal new
fundamental physics. 
For example, a proven inconsistency between inferences at
high redshift from the study of the cosmic microwave background, with
inferences at lower redshift from galaxy redshift surveys would
challenge the standard description of the evolution of the Universe
over this redshift interval, and possibly lead to revisions of either
our theory of gravity or of our assumptions about the nature of dark
matter and dark energy. 

In this paper we present new results from an 
observational program aimed at
precision cosmography using gravitational lens time delays. 
The idea of doing
cosmography with time-delay lenses goes back fifty years and
it is a simple one \citep{Refsdal64}. When a source is observed
through a strong gravitational lens, multiple images form at the
extrema of the time-delay surface, according to Fermat's principle
\citep[e.g.,][]{SEF92, Falco05,SchneiderEtal06}. If the
source is variable, the time delays between the  
images can be measured by careful monitoring of the image light
curves
\citep[see, \eg][]{Courbin03}. 
With an
accurate model of the gravitational lens, the absolute time delays can
be used to convert angles on the sky into an absolute distance, 
the so-called time-delay distance,
which can be compared with predictions from the cosmological
model given the lens and source redshifts \citep[\eg][and
references therein]{BlandfordNarayan92, Jackson07, Treu10}.
This distance is a combination of three angular diameter 
distances, and so is primarily sensitive to the Hubble constant ($H_0$), 
with some higher order dependence on the other cosmological parameters
\citep{CoeMoustakas09,Linder11}. Gravitational
time delays are a one-step cosmological method to determine the Hubble
constant that is completely independent of the local cosmic distance
ladder \citep{FreedmanEtal01,RiessEtal11, FreedmanEtal12, ReidEtal12}.  
Knowledge of
the Hubble constant is currently the key limiting factor in measuring
parameters like the dark energy equation of state, curvature, or
neutrino mass, in combination with other probes like the cosmic
microwave background \citep{FreedmanMadore10,RiessEtal11, FreedmanEtal12,
WeinbergEtal12, SuyuEtal12b}.  These features make strong gravitational
time delays a very attractive probe of cosmology.

Like most high-precision measurements, however, a good idea is only
the starting point. A substantial amount of effort and observational
resources needs to be invested to control the systematic errors. In the
case of gravitational time delays, this has required several observational
and modeling breakthroughs. Accurate, long duration, and 
well-sampled light
curves are necessary to obtain accurate time delays in the presence
of microlensing. Modern light curves have much higher photometric
precision, sampling and duration
\citep{FassnachtEtal02,CourbinEtal11} compared to the early 
pioneering light 
curves \citep[\eg][]{LeharEtal92}. High resolution images of
extended features in the source, and stellar kinematics of the main
deflector, provide hundreds to thousands of data points to constrain
the mass model of the main deflector, thus reducing the degeneracy
between the distance and the gravitational potential of the lens
that affected previous
models constrained only by the positions of the lensed quasars
\citep[e.g.,][]{SchechterEtal97}. Finally, cosmological numerical
simulations can now be used to characterize the distribution of mass along
the line of sight (LOS)
\citep{HilbertEtal09}, which was usually neglected in early studies
that were not aiming for precisions of a few percent.  
The advances in the use
of gravitational time delays as a cosmographic probe are summarized
in the analysis of the gravitational lens system \blens\ by
\citet{SuyuEtal10}. 
In that paper, we demonstrated that, with sufficient
ancillary data, a single gravitational lens can yield a time-delay
distance measured to 5\% precision, and the Hubble constant to 7\%
precision. In combination with the \textit{Wilkinson
  Microwave Anisotropy Probe} 5-year (WMAP5) results, 
the \blens\ time-delay distance
constrained $w$ to 18\% precision and the curvature parameter to
$\pm0.02$ precision, comparable to contemporary Baryon Acoustic
Oscillation experiments \citep{PercivalEtal07} and observations of the
growth of massive galaxy clusters \citep[on $w$
constraints;][]{MantzEtal10}.

Building on these recent developments in the analysis, and on
the state-of-the-art monitoring campaigns carried out by the
COSMOGRAIL (COSmological MOnitoring of GRAvItational Lenses; e.g.,
\citeauthor{VuissozEtal08}~\citeyear{VuissozEtal08}; 
\citeauthor{CourbinEtal11}~\citeyear{CourbinEtal11}; 
\citeauthor{TewesEtal12b}~\citeyear{TewesEtal12b}) and
\citet{KochanekEtal06} teams,
it is now possible to take gravitational time delay lens
cosmography to the next level and achieve precision comparable to
current measurements of the Hubble constant, flatness, $w$ and other
cosmological parameters \citep{RiessEtal11, FreedmanEtal12, KomatsuEtal11}.  
To this end we have recently initiated a program to obtain data and
model four additional gravitational lens systems with the same quality
as that of \blens.  These four lenses are selected from the
  COSMOGRAIL sample with the smallest uncertainties in the delays
  between the images of $\leq$$6\%$.  They cover various image
  configurations: (1) four lensed images with three of them merging,
  a.k.a.~the ``cusp'' configuration, (2) four lensed images with two
  of them merging, a.k.a.~the ``fold'' configuration, (3) four images
  that are nearly symmetric about the lens center, a.k.a.~the ``cross''
  configuration, and (4) two images on opposite sides of the lens
  galaxy.  This sample will allow us to probe the optimal lens
  configuration for time-delay cosmography and also investigate
  potential selection effects.

We present here the results for the first of these systems, \rxj,
based on new time delays measured by the COSMOGRAIL collaboration
\citep{TewesEtal12b}, new spectroscopic data from the
Keck Telescope, a new analysis of archival \textit{Hubble Space Telescope} (\hst)
images, and a characterization of the LOS effects through
numerical simulations, the observed galaxy number counts in the field, and 
the modeled external shear. We carry out a self-consistent modeling of
all the available data sets in a Bayesian framework, and infer
(1) a likelihood function
for the time-delay distance that can be combined with any other 
independent probe of cosmology, 
and (2) in combination with our previous measurement of \blens\ and the
WMAP 7-year (WMAP7) results, the posterior probability density  function (PDF) for the
Hubble constant, curvature density parameter and dark energy equation-of-state
parameter $w$.

Three additional lens systems are scheduled to be observed with \hst\ in cycle
20 (GO 12889; PI Suyu) and will be published in forthcoming papers.
An integral part of this program is the use of blind analysis, 
to uncover unknown systematic errors and to avoid unconscious 
experimenter bias.
Only when each system's analysis has been judged to be
complete and final by its authors, 
are the implications for cosmology
revealed. These results are then published without any further
modification. In this way, 
we can assess whether the results are mutually
consistent within the estimated errors or whether unknown systematics
are adding significantly to the total error budget.

This paper is organized as follows. After a brief recap of the theory
behind time-delay lens cosmography in 
\sref{sec:theory}, we summarize our strategy in
\sref{sec:strgy} and describe our observational data in 
\sref{sec:obs}. In
\sref{sec:prob}, we write out the probability theory used in the
data modeling and describe the procedure for carrying out the blind
analysis.  The lensing and time-delay analysis are presented in 
\sref{sec:lensmod}, and a description of our treatment of the
LOS mass structure in the \rxj\ field is in 
\sref{sec:breakmsd}.   We present measurements of the time-delay
distance, and discuss the sources of uncertainties in
\sref{sec:tdist}.  We show our unblinded  
cosmological parameter inferences in
\sref{sec:cosmo}, which includes joint analysis with our
previous lens data set and with WMAP7. Finally, we conclude in
\sref{sec:conclude}.  Throughout this paper, each quoted parameter
estimate is the median of the appropriate one-dimensional marginalized
posterior PDF, with the quoted uncertainties showing, unless otherwise
stated, the $16^{\rm th}$ and $84^{\rm th}$ percentiles (that is, the
bounds of a 68\% credible interval).


\section{Cosmography from gravitational lens time delays}
\label{sec:theory}

In this section, we give a brief overview of the use of gravitational
lens time delays to study cosmology.  More details on the subject can
be found in, e.g., \citet{SchneiderEtal06},
\citet{Jackson07}, \citet{Treu10} and \citet{SuyuEtal10}.  
Readers familiar with time-delay lenses may wish to proceed directly
to \sref{sec:strgy}.

In a gravitational lens system, the time it takes the light from the
source to reach us depends on both the path of the light ray and also
the gravitational potential of the lens.  The excess time delay of an
image at angular position $\imVec=(\im_1,\im_2)$ with corresponding
source position $\srVec=(\sr_1,\sr_2)$ relative to the case of no
lensing is
\be
t(\imVec, \srVec) = \frac{\tdist}{c} \left[ \frac{(\imVec-\srVec)^2}{2}-\psi(\imVec) \right],
\ee
where $\tdist$ is the so-called time-delay distance, $c$ is the speed
of light, and $\psi(\imVec)$ is the lens potential.  The time-delay
distance is a combination of the angular diameter distance to the
lens (or deflector) ($D_{\rm d}$) at redshift $\zd$, to the source
($D_{\rm s}$), and between the lens and the source ($D_{\rm ds}$):
\be
\label{eq:tdist}
\tdist \equiv (1+\zd) \frac{D_{\rm d} D_{\rm s}}{D_{\rm ds}}.
\ee
The lens potential $\psi(\imVec)$ is related to the dimensionless
surface mass density of the lens, $\kappa(\imVec)$, via
\be
\label{eq:poisson}
\nabla^2 \psi(\imVec) = 2 \kappa(\imVec),
\ee
where 
\be
\label{eq:kappadef}
\kappa(\imVec) = \frac{\Sigma(D_{\rm d}\imVec)}{\Sigma_{\rm crit}},
\ee
$\Sigma(D_{\rm d}\imVec)$ is the surface mass density of the lens
(the projection of the three dimensional density $\rho$ along the
LOS), $\Sigma_{\rm crit}$ is the critical surface mass density 
defined by
\be
\label{eq:scrit}
\Sigma_{\rm crit} = \frac{c^2}{4\pi G}\frac{D_{\rm s}}{D_{\rm d}
  D_{\rm ds}},
\ee
and $G$ is the gravitational constant.

For lens systems whose sources vary in time (as do active galactic
nuclei, AGNs), one can monitor the brightnesses of the lensed images
over time and hence measure the time delay, $\Delta t_{ij}$, between
the images at positions $\imVec_i$ and $\imVec_j$:
\bea
\nonumber \Delta t_{ij} &\equiv& t(\imVec_i,\srVec) -
t(\imVec_j,\srVec) \\
\label{eq:dt}
&=& 
\frac{\tdist}{c} \left [  \frac{(\imVec_i-\srVec)^2}{2}-\psi(\imVec_i)
  - \frac{(\imVec_j-\srVec)^2}{2}+\psi(\imVec_j) \right ]. 
\eea
By using the image configuration and morphology, one can model the
mass distribution of the lens to determine the lens potential
$\psi(\imVec)$ and the unlensed source position $\srVec$.  Lens
systems with time delays can therefore be used to measure $\tdist$ via
\eref{eq:dt}, and constrain cosmological models via the
distance-redshift test
\citep[\eg][]{Refsdal64, Refsdal66, FadelyEtal10, SuyuEtal10}.
Having dimensions of distance,
$\tdist$ is inversely proportional to $H_0$, and
being a combination of three angular diameter distances, it
depends weakly on the other cosmological parameters as well.

The radial slope of the lens mass distribution and the time-delay
distance both have direct influence on the observables: 
for a given time delay, a galaxy with
a steep radial profile leads to a lower $\tdist$ than that of a galaxy
with a shallow profile \citep[\eg][]{WittEtal00, Wucknitz02,Kochanek02}.
Therefore, to measure $\tdist$, it is necessary to
determine the radial slope of the lens galaxy.  
Several authors have shown that the spatially extended
sources (such as the host galaxy of the AGN in time-delay lenses) can 
be used to measure the radial slope at the image positions,
where it matters
\citep[\eg][]{DyeWarren05, DyeEtal08, SuyuEtal10, VegettiEtal10,
Suyu12}.

In addition to the mass distribution associated with the lens galaxy,
structures along the LOS also affect the 
observed time
delays.  The external masses and voids cause additional focussing and
defocussing of the light rays respectively, and therefore affect the
time delays and $\tdist$ inferences.  
We follow \citet{Keeton03}, \citet{SuyuEtal10} and many others and suppose
that the effect of the LOS
structures can be characterized by a single parameter, the 
external convergence 
$\kext$, with positive values associated with overdense LOS and negative
values with underdense LOS.  Except for galaxies very nearby
to the strong
lens system, the $\kext$ contribution of the LOS structures to the
lens is effectively constant across the scale of the lens system.

Given the measured delays between the images of a strong lens, a mass
model that does not account for the external
convergence leads to an under/overprediction of $\tdist$ for
over/underdense LOS.  In particular, the true $\tdist$ is related to
the modeled one by 
\be
\label{eq:msd:tdist}
\tdist = \frac{\tdistmod}{1-\kext}.
\ee
Two practical approaches to
overcome this degeneracy are 
(1) to use the stellar kinematics of the lens galaxy
\citep[\eg][]{TreuKoopmans02,KoopmansTreu03, TreuKoopmans04, BarnabeEtal09,
AugerEtal10, SuyuEtal10, SonnenfeldEtal12} to make an independent
estimate of the lens mass
and (2) to study the
environment of the lens system \citep[\eg][]{KeetonZabludoff04,
FassnachtEtal06, MomchevaEtal06, SuyuEtal10, WongEtal11,
FassnachtEtal11} in order to estimate $\kext$ directly.
In \sref{sec:breakmsd}, we combine both
approaches to infer $\kext$.


\section{Accurate and precise distance measurements}
\label{sec:strgy}

We summarize our strategy for accurate and precise cosmography with
all known sources of systematic uncertainty taken into account.  We
assemble the following key ingredients for obtaining $\tdist$ via
\eref{eq:dt}
\begin{itemize}
\item observed time delays: dedicated and long-duration monitoring,
  particularly from COSMOGRAIL, yields delays with uncertainties of
  only a few percent \citep{TewesEtal12a,TewesEtal12b}.
\item lens mass model: deep and high-resolution imagings of the lensed
  arcs, together with our flexible modeling techniques that use as
  data the thousands of surface brightness pixels of the lensed
  source, allow constraints of the potential difference between the
  lensed images (in \eref{eq:dt}) at the few percent level
  \citep[e.g.,][]{SuyuEtal10}. 
\item external convergence: the stellar velocity dispersion of the
  lens galaxy provides constraints on both the lens mass distribution
  and external convergence.  We further calibrate observations of
  galaxy counts in the fields of lenses \citep{FassnachtEtal11} with
  ray tracing through numerical simulations of large-scale structure
  \citep[\eg][]{HilbertEtal07} to constrain directly and statistically
  $\kext$ at the $\sim$5\% level \citep{SuyuEtal10}.
\end{itemize}
With all these data sets for the time-delay lenses, we can measure
$\tdist$ for each lens with $\sim 5-8\%$ precision (including all
sources of known uncertainty).  A comparison of a sample of lenses
will allow us to test for residual systematic effects, if they are
present.  With systematics under control, we can combine the
individual distance measurements to infer global properties of
cosmology since the gravitational lenses are independent of one
another.


\section{Observations of \rxj}
\label{sec:obs}

The gravitational lens \rxj\
(J2000: ${\rm 11^{h}31^{m}52^{s}}$, $-12^{\degr}31'59''$) was discovered by
\citet{SluseEtal03} 
during polarimetric imaging of a sample of radio quasars.  The
spectroscopic redshifts of the lens and the quasar source are
$\zd=0.295$ and $\zs=0.658$, respectively \citep{SluseEtal03}.  We present the archival
\hst\ images in 
\sref{sec:obs:hst}, the time delays from COSMOGRAIL in
\sref{sec:obs:td}, the lens velocity dispersion in 
\sref{sec:obs:vd} and
information on the lens environment in \sref{sec:obs:env}.

\subsection{Archival \hst\ imaging}
\label{sec:obs:hst}
\begin{figure}
  \centering
  \includegraphics[width=0.45\textwidth]{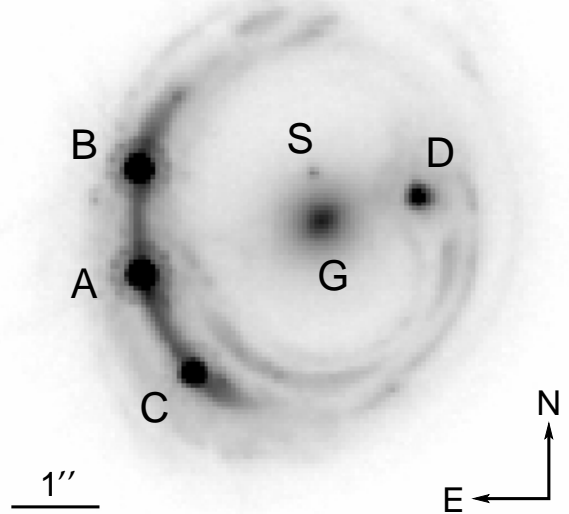}
  \caption{\label{fig:hst} \hst\ ACS F814W image of the gravitational
    lens \rxj.  The lensed AGN images of the spiral source galaxy are
    marked by A, B, C and D, and the star forming regions of the
    spiral galaxy form the spectacular lensed structures.  The 
    primary lens galaxy and the satellite lens galaxy are indicated by G
    and S, respectively.}
\end{figure}

\hst Advanced Camera for Surveys (ACS) images were obtained for \rxj\
in two filters, F814W and F555W (ID 9744; PI Kochanek).  In each
filter, five exposures were taken with a total exposure time of 1980s.
We show in~\fref{fig:hst} the F814W image of the lens system.  The
background quasar source is lensed into four images denoted by A, B,
C, and D, and the spectacular features surrounding the quasar images
are the lensed images of the quasar host that is a spiral
galaxy \citep{ClaeskensEtal06}.  The primary lens galaxy is marked by
G, and the object marked by S is most likely a satellite of G
\citep{ClaeskensEtal06}.  Henceforth, we refer to S as the satellite.

We reduce the images using {\sc MultiDrizzle}\footnote{{\sc MultiDrizzle} is a product of the Space Telescope Science Institute, which is operated by AURA for NASA.} with charge transfer inefficiency taken into account \citep[\eg][]{AndersonBedin10,MasseyEtal10}. The images are drizzled to a final pixel scale of $0\farcs05$~pix$^{-1}$ and the uncertainty on the flux in each pixel is estimated from the science and the weight image by adding in quadrature the Poisson noise from the source and the background noise due to the sky and detector readout. We note that in some of the exposures the central regions of the two brightest AGN images are slightly saturated and are masked during the drizzling process.

To model the lens system using the spatially extended Einstein ring of
the host galaxy, we focus on the F814W image since the contrast
between the ring and the AGN is more favorable in F814W.  In
particular, the bright AGNs in F555W have diffraction spikes extending
into the Einstein ring that are difficult to model and are thus prone
to systematics effects.  
The detailed modeling of the F814W image is in \sref{sec:lensmod}.

\subsection{Time delays}
\label{sec:obs:td}

We use the new time-delay measurements of \rxj\ presented in
\citet{TewesEtal12b}. 
The COSMOGRAIL and Kochanek et al.~teams have monitored \rxj\ since
December 2003 using several optical 1--1.5m telescopes.
Resolved light curves of the four AGN images are extracted from these observations by ``deconvolution photometry'', following \citet{MagainEtal98}.
These curves presently span 9 years with over 700 epochs, and display a typical sampling of 2--3 days within the observation seasons.
The time delays are measured through several new and independent techniques \citep[detailed in][]{TewesEtal12a}, all specifically developed to handle microlensing variability due to stars in the lens galaxy. 
All these techniques yield consistent results, attributed to both the long light curves and the comprehensive uncertainty estimation.
For our analysis we select the time-delay measurements from the
\emph{regression difference technique} as recommended by
  \citet{TewesEtal12b} who showed that this technique yielded the
  smallest bias and variance in their error analysis when applied to
  synthetic curves mimicking the microlensing variability in \rxj.
In particular we use the time delays relative to image B, namely:
$\dt_{\rm AB}=0.7 \pm 1.4$ days, $\dt_{\rm CB}= -0.4 \pm 2.0$ days,
and $\dt_{\rm DB}=91.4\pm 1.5$ days, where the uncertainties are
conservative and direct sums of the estimated statistical and
systematic contributions from \citet{TewesEtal12b}. 

\subsection{Lens velocity dispersion}
\label{sec:obs:vd}

We observed \rxj\ with the Low-Resolution Imaging Spectrometer
(LRIS; \citeauthor{OkeEtal95}~\citeyear{OkeEtal95}) on Keck 1 on 4--5
January 2011. The data were obtained from the red side of the
spectrograph using the 600/7500 grating with the D500 dichroic in
place.  A slit mask was employed to obtain simultaneously spectra for
galaxies near the lens system. The night was clear with a nominal
seeing of $0\farcs7$, and we use 4 exposures of 1200s for a total
exposure time of 4800s.  

We follow \citet{AugerEtal08} to reduce each exposure by performing a
single resampling of the spectra onto a constant wavelength grid.  We
use the same wavelength grid for all exposures to avoid resampling the
spectra when combining them.  An output pixel scale of 0.8 \AA\ pix$^{-1}$ was
used to match the dispersion of the 600/7500 grating.  Individual
spectra are extracted from an aperture 0\farcs81 wide (corresponding
to 4 pixels on the LRIS red side) centered on the lens galaxy.  The
size of the aperture was chosen to avoid contamination from the
spectrum of the lensed AGNs.  We combine the extracted spectra by
clipping the extreme points at each wavelength and taking the
variance-weighted sum of the remaining data points. We repeat the same
extraction and coaddition scheme for a sky aperture to determine the
resolution of the output co-added spectrum: ${\rm R} = 2300$,
corresponding to $\sigma_{\rm obs} = 56\, {\rm km\, s^{-1}}$. The
typical signal-to-noise ratio per pixel of the final spectrum is $\sim$$20$.

\begin{figure}
\centering\includegraphics[width=75mm,clip]{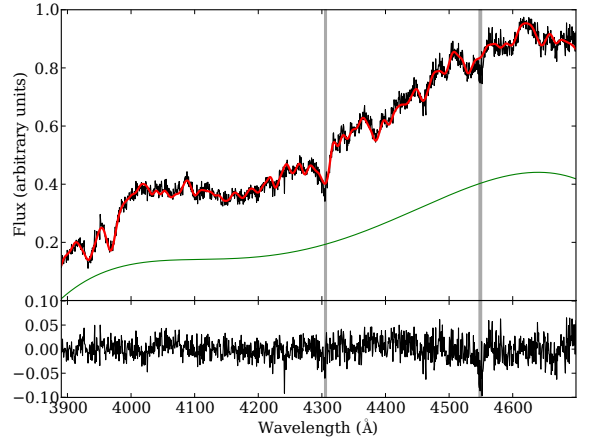}
  \caption{\label{fig:LRISvelocitydispersion} Top panel: The LRIS
    spectrum of \rxj\ (black line) with a model generated from 9
    INDO-US templates and a 5$^{\rm th}$ order continuum overplotted (red
    line, with green showing the continuum). The gray shaded areas
    were not included in the fit.
    Bottom panel: the residuals of the model fit. From the spectrum
    and model, we measure a central velocity dispersion of $\sigma = 323
    \pm 20\, {\rm km\, s^{-1}}$, including systematic uncertainties. }
\end{figure}

The stellar velocity dispersion is determined in the same manner as
\citet{SuyuEtal10}. Briefly, we use a suite of stellar templates of
K and G giants, augmented with one A and one F star template, from the
INDO-US library \citep{ValdesEtal04} to fit directly to the observed
spectrum, after convolving each template with a kernel to bring them to
the same spectral resolution as the data. First and second velocity moments
are proposed by a Markov chain Monte Carlo (MCMC) simulation and the templates
are shifted and broadened to these moments. We then fit the model
templates to the data in a linear least squares sense, including a fifth
order polynomial to account for any emission from the background source
\citep[\eg][]{SuyuEtal10}. The observed and modeled spectra are
shown in \fref{fig:LRISvelocitydispersion}. Our estimate for the
central line-of-sight velocity
dispersion from this inference is $\sigma = 323 \pm 20\, {\rm km\, s^{-1}}$,
including systematics from changing the polynomial order and choosing
different fitting regions.

\subsection{Galaxy counts in the field}
\label{sec:obs:env}

\citet{FassnachtEtal11} counted the number of galaxies with F814W
magnitudes between 18.5 and 24.5 that lie within $45''$ from the lens
system.  Compared to the aperture counts in random lines of sight in
pure-parallel fields\footnote{The ``pure-parallel fields''
    consist of 20 points in the ACS F814W filter selected from a
    pure-parallel program that searched for emission line galaxies in
    random fields (GO 9468; PI Yan).},
\rxj\ has 1.4 times the average number of
galaxy counts \citep{FassnachtEtal11}.  We use this relative galaxy count in 
\sref{sec:breakmsd:env} to estimate statistically the external
convergence.


\section{Probability theory for combining multiple data sets}
\label{sec:prob}

We now present the mathematical framework for the
inference of cosmological parameters from the combination of 
the data sets described in the previous section.  In
order to test for the presence of any unknown systematic uncertainty,
we describe a procedure for blinding the results during the analysis
phase in \sref{sec:prob:blind}. This procedure is designed to ensure
against unconscious experimenter bias towards ``acceptable'' results.


\subsection{Joint analysis}
\label{sec:prob:bayes}

The analysis performed here is similar to the one presented in
\citet{SuyuEtal10}, with a few improvements.  We briefly describe the
procedure below.

The data sets are denoted by $\dacsVec$ for the ACS image (packaged into 
a vector of $160^2$ surface brightness values), $\dtVec$ for the delays between the
images, $\vdisp$ for the lens velocity dispersion, and $\denvVec$ for
properties of the lens environment such as the relative galaxy count
$n_{\rm r}=n_{\rm gal}/\langle n_{\rm gal}\rangle$.  We are interested
in obtaining the posterior PDF of the model
parameters $\parAllVec$ given all available data,
\be
\label{eq:Pall}
P(\parAllVec | \dacsVec, \dtVec, \vdisp, \denvVec) \propto
P(\dacsVec,\dtVec,\vdisp,\denvVec|\parAllVec) P(\parAllVec)
\ee
where the proportionality follows from Bayes' Theorem.  The first term
on the right-hand side 
is known as the likelihood, and the second is the prior PDF.  
Since the data sets are independent, the likelihood is separable,
\bea
\label{eq:LikeSep}
\nonumber P(\dacsVec,\dtVec,\vdisp,\denvVec|\parAllVec) & = &
P(\dacsVec|\parAllVec)\,P(\dtVec|\parAllVec) \\
& & \ \ \ \ P(\vdisp|\parAllVec)\,P(\denvVec|\parAllVec).
\eea

Some of the parameters influence all the predicted data sets, while 
other parameters affect the fitting of particular data sets only.
Specifically, $\parAllVec = \{\parCosVec, \slope, \thEin,
\gext, \parLensVec, \rani, \kext\}$, where $\parCosVec$ are the
cosmological parameters (e.g., $H_0$, $w$, $\Ode$), $\slope$ is the
radial profile slope of the main lens galaxy (where $\rho \propto
r^{-\slope}$), $\thEin$ is the Einstein radius of the main lens (that
characterizes the normalization of the lens mass profile),
$\gext$ is the external shear strength at the lens, $\parLensVec$
denotes the remaining lens model parameters for the ACS 
data,\footnote{excluding
the source surface brightness parameters $\esrVec$ that can be
marginalized analytically} $\rani$ is the anisotropy
radius for the stellar orbits of the lens galaxy, and $\kext$ is the
external convergence.  For the lensing and time delays, we subsume the
cosmological dependence into the time-delay distance 
$\tdist = \tdist(\parCosVec)$.  Keeping
only the direct dependencies in each of the likelihoods, we obtain
\bea
\label{eq:postParAll}
\nonumber & P(\parAllVec | \dacsVec, \dtVec, \vdisp, \denvVec) \propto P(\dacsVec, \dtVec
| \tdist, \slope, \thEin, \gext,\\
\nonumber & \parLensVec, \kext) P(\vdisp | \parCosVec,
\slope, \thEin, \rani, \kext) P(\denvVec|\kext,\gext) \\ 
& P(\parCosVec) P (\slope) P(\thEin) P(\gext) P(\parLensVec) P(\rani) P(\kext).
\eea

For cosmography, we are interested in the cosmological parameters
$\parCosVec$ after marginalizing over all other parameters
\bea
\label{eq:postCosmo}
\nonumber & P(\parCosVec|\dacsVec, \dtVec, \vdisp, \denvVec) = \int
{\rm d}\slope\,{\rm d}\thEin\,{\rm d}\gext\,{\rm d}\parLensVec\,{\rm
  d}\rani\,\\
&{\rm d}\kext\, P(\parAllVec | \dacsVec, \dtVec, \vdisp, \denvVec) 
\eea
We describe the forms of the partially marginalized lensing and
time-delay likelihood in \sref{sec:lensmod}, the kinematics
likelihood in \sref{sec:breakmsd:vd} and the external convergence
likelihood in \sref{sec:breakmsd:env}.  For marginalizing the
parameters that are common to the data sets, we importance sample the
priors following \citet{LewisBridle02} and \citet{SuyuEtal10} 
(a procedure sometimes referred to as ``simple Monte Carlo'').


\subsection{Blind analysis}
\label{sec:prob:blind}

We blind the analysis to avoid experimenter bias, allowing us to  test
for the presence of residual systematics in our analysis technique by
comparing the final unblinded results from \rxj\ with the constraints from the
previous analysis of \blens.  As described by \citet{ConleyEtal06}, the 
blinding is not meant to hide all information from the experimenter; rather,
we blind only the parameters that concern the cosmological inference.

We define two analysis phases. During the initial ``blind'' phase, we 
compute likelihoods and priors, and sample the posterior PDF, as given above, 
taking care to only make parameter-space plots using one plotting code. This
piece of software  adds offsets to the cosmological parameters ($\tdist$ and
the components of $\parCosVec$) before displaying the PDFs, such that we
always see the marginalized distributions with centroids at exactly zero. We
can therefore still see, and measure, the {\it precision} of the blinded 
parameters, and visualize the correlations between these parameters, but
without being able to see if we have ``the right answer'' based on
our expectations.  Both the parameter
uncertainties and degeneracies serve as useful checks during this blind phase:
the plotting routine can overlay the constraints from different models to
investigate sources of statistical and systematic uncertainties.

During the blind phase we performed a number of tests on the modeling to
quantify the sources of uncertainties, and to check the robustness of the
results.  These are described in Sections \ref{sec:obs}--\ref{sec:tdist}. At
the end of the tests, the collaboration convened a telecon to unblind the
results.  The authors SHS, MWA, SH, PJM, MT, TT, CDF, LVEK, DS, and FC
discussed in detail the analysis and the blinded results,  over a summary
website.  After all agreeing that the blind analysis was complete, and that we
would publish without modification the results once unblinded, a script was
run to update automatically the same website with plots and tables containing 
cosmological constraints no longer offset to zero. These are the results 
presented in Sections \ref{sec:tdist} and \ref{sec:cosmo:rxj}.


\section{Lens Modeling}
\label{sec:lensmod}

In this section, we simultaneously model the ACS images and the time
delays to measure the lens model parameters, particularly $\tdist$,
$\slope$, $\thEin$, and $\gext$.


\subsection{A comprehensive mass and light model}
\label{sec:lensmod:lenssr}

The ACS image in \fref{fig:hst} shows the light from the source
as lensed by the galaxies G and S.  To predict the surface brightness
of the pixels on the image, we need a model for the lens
mass distribution (that deflects the light of the source), the lens
light distribution, the source light distribution and the point spread
function (PSF) of the telescope.  


\subsubsection{Lens mass profiles}
\label{sec:lensmod:lenssr:lensmass}

We use elliptically-symmetric distributions with 
power-law profiles to model the dimensionless surface mass
density of the lens galaxies,
\be
\label{eq:PLkappa}
\kappa_{\rm pl}(\im_1,\im_2) =
\frac{3-\slope}{2}\left(\frac{\thEin}{\sqrt{q \im_1^2+\im_2^2/q }} \right)^{\slope-1},
\ee
where $\slope$ is the radial power-law slope (with $\slope=2$
corresponding to isothermal), $\thEin$ is the Einstein radius, and $q$
is the axis ratio of the elliptical isodensity contours.  Various
studies have shown that the power-law profile provides accurate
descriptions of lens galaxies \citep[\eg][]{GavazziEtal07,
HumphreyBuote10, KoopmansEtal09, AugerEtal10, BarnabeEtal11}.  In
particular, \citet{SuyuEtal09} found that the grid-based lens
potential corrections from power-law models were only $\sim 2\%$ for
\blens\ with interacting lens galaxies, thus validating the use of the
simple power-law models even for complicated lenses. We note that the
surface brightness of the main deflector in \rxj\ shows no signs of
interaction (Section \ref{sec:lensmod:lenssr:lenslight}) and it is therefore
much simpler than the case of \blens,  further justifying the use
of a simple power-law model to describe the mass distribution within the
multiple images.

The Einstein radius in \eref{eq:PLkappa} corresponds to the geometric
radius of the critical curve,\footnote{The critical curve of
  $\kappa_{\rm pl}$ in \eref{eq:PLkappa} is symmetric
  about $\im_1$ and $\im_2$, and the geometric radius is $\sqrt{\theta_{\rm
      long} \theta_{\rm short}}$, where $\theta_{\rm long}$ ($\theta_{\rm short}$) is
  the distance of the furthest (closest) point on the critical curve from
  the origin.} and the mass enclosed within the 
isodensity contour with the geometric Einstein radius is 
\be
\label{eq:MEin}
M_{\rm E} = \pi \thEin^2 D_{\rm d}^2 \Sigma_{\rm crit}
\ee
that depends only on $\thEin$, a robust quantity in lensing.  

The deflection angle and lens potential of the power-law profile are
computed following \citet{Barkana98}. For each lens galaxy, the
distribution is suitably translated to the position of the lens galaxy
and rotated by the position angle $\phi$ of the lens galaxy (where
$\phi$ is a free parameter,
measured counterclockwise from $\im_2$).  Since the
satellite galaxy is small in extent, we approximate its mass
distribution as a spherical isothermal mass distribution with
$\slope_{\rm S}=2$ and $q_{\rm S}=1$ in 
\eref{eq:PLkappa}.  The (very small) impact of the satellite on cosmographic
inferences is discussed in \sref{sec:sourceError}.

Our coordinate system is defined such that $\im_1$ and $\im_2$ point
to the west and north, respectively.  The origin of the coordinates is
at the bottom-left corner of the ACS image containing 160$\times$160
pixels.

In addition to the lens galaxies, we include a
constant external shear of the following form in polar coordinates
$\im$ and $\varphi$: 
\be
\label{eq:extsh}
\psi_{\rm ext}(\im,\varphi) = \frac{1}{2}\gamma_{\rm ext} \im^2
\cos2(\varphi - \phi_{\rm ext}),
\ee
where $\gamma_{\rm ext}$ is the shear strength and $\phi_{\rm ext}$ is
the shear angle.  The shear position angle of $\phi_{\rm ext}=0\degr$
corresponds to a shearing along the $\im_1$ direction whereas
$\phi_{\rm ext}=90\degr$ corresponds to a shearing in the $\im_2$
direction.  

We do not include the external convergence $\kext$ at this stage, 
since this parameter is
completely degenerate with $\tdist$ in the ACS and time-delay
modeling.  Rather, we use $\tdistmod \equiv (1-\kext)\tdist$ for the
lensing and time-delay data, and information on $\kext$ will come from
kinematics and lens environment in \sref{sec:breakmsd} to allow
us to infer $\tdist$.

\subsubsection{Lens light}
\label{sec:lensmod:lenssr:lenslight}

For the light distribution of the lens galaxies, we use elliptical
S{\'e}rsic profiles,
\be
\label{eq:sersic}
I(\theta_1,\theta_2) = A \exp \left[ -k \left (\left(\frac{\sqrt{\theta_1^2 +
      \theta_2^2/q_{\rm L}^2}}{R_{\rm eff}}\right) ^ {1/n_{\rm
    sersic}} - 1 \right) \right],
\ee
where $A$ is the amplitude, $k$ is a constant such that $R_{\rm eff}$
is the effective (half-light) radius, $q_{\rm L}$ is the axis ratio, and
$n_{\rm sersic}$ is the S{\'e}rsic index \citep{Sersic68}.  The
distribution is suitably rotated by positions angle $\phi_{\rm L}$ and
translated to the galaxy positions $(\theta_{1,L}, \theta_{2,L})$.  
We find that a single S{\'e}rsic profile for the 
primary lens galaxy leads to significant residuals, as was found by
\citet{ClaeskensEtal06}. Instead, we use two S{\'e}rsic profiles
with common centroids and position angles to describe the lens galaxy G.
For the small satellite galaxy that illuminates only a few pixels, we
use a circular S{\'e}rsic profile with $n_{\rm sersic}=1$.  This
simplifying assumption has no effect on the mass modeling since the
light of the satellite is central and compact, and thus does not
affect the light or mass of the other components.

\subsubsection{Source light}
\label{sec:lensmod:lenssr:srlight}

To describe the surface brightness distribution of the lensed source,
we follow \citet{Suyu12} and use a hybrid model comprised of (1) point
images for the lensed AGNs on the image plane, and (2) a regular grid
of source surface brightness pixels for the spatially extended
AGN host galaxy.  Modeling the AGN point images independently accommodates
variations in the fluxes arising from microlensing, time delays
and substructures.  Each AGN image therefore has three parameters: a
position in $\im_1$ and $\im_2$ and an amplitude.  We collectively
denote these AGN parameters as $\agnVec$.  The extended source
on a grid is modeled following \citet{SuyuEtal06}, with curvature
regularization.

\subsubsection{PSF}
\label{sec:lensmod:lenssr:psf}

A PSF is needed to model the light of the lens galaxies and the lensed
source.  We use stars in the field to approximate the PSF, which has
been shown to work sufficiently well in modeling galaxy-scale strong lenses
\citep[\eg][]{MarshallEtal07, SuyuEtal09, Suyu12}.  In particular,
we adopt the star that is located at $2\farcm4$ northwest of the lens
system as the model of the PSF.

\subsubsection{Image pixel uncertainties}
\label{sec:lensmod:lenssr:noise}

The comprehensive mass and light model described above
captures the large-scale features
of the data very well.  However, small-scale features in the image
might cause misfits which, if not taken into
account, may lead to an underestimation of parameter uncertainties and
biased parameter estimates.  \citet{Suyu12} found that by boosting
the pixel uncertainty of the image surface brightness, the lens model
parameters can be faithfully recovered with realistic estimation of
uncertainties.

Following this study, we introduce two terms to describe the
variance of the intensity at pixel $i$ of the ACS image $\dacsVec$,
\be
\label{eq:noiseboost}
\sigma_{{\rm pix},i}^2 = \sigma_{\rm bkgd}^2 + f \dacsi ,
\ee
where $\sigma_{\rm bkgd}$ is the background uncertainty, $f$ is a
scaling factor, and $\dacsi$ is the image intensity.  The second term,
$f \dacsi$, corresponds to a scaled version of Poisson noise for the
astrophysical sources.  We measure $\sigma_{\rm bkgd}$ from a blank
region in the image without astrophysical sources.  We set the value
of $f$ such that the reduced $\chi^2$ is $\sim$1 for the lensed image
reconstruction \citep[see, e.g.,][for details on the computation of
the reduced $\chi^2$ that takes into account the regularization on the
source pixels]{SuyuEtal06}. \eref{eq:noiseboost} by
design downweights 
regions of high intensities where the residuals are typically most
prominent.  This allows the lens model to fit to the overall structure
of the data instead of reducing high residuals at a few locations at the
expense of poorer fits to the large-scale lensing features.  The
residuals near the AGN image positions are particularly high due to
the high intensities and slight saturations in some of the images.
Therefore, we set the uncertainty on the inner pixels of the AGN
images to a very large number that effectively leads to these pixels being
discarded.
We discard only a small region in fitting the AGN light, and increase
the region to minimize AGN residuals when using the extended source
features to constrain the lens mass parameters.


\subsection{Likelihoods}
\label{sec:lensmod:like}

The model-predicted image pixel
surface brightness can be written as a vector 
\be
\label{eq:dataP}
\dacsVecP =  \Bmat \galVec + \Bmat\Lmat \esrVec +
\sum_{i=1}^{N_{\rm AGN}} \eagnVec_i(\agnVec),
\ee
where $\Bmat$ is a blurring operator that accounts for the PSF convolution,
$\galVec$ is a vector of image pixel intensities of the S{\'e}rsic
profiles for the lens galaxy light, $\Lmat$ is the lensing operator
that maps source intensity to the image plane based on the 
deflection angles computed from the parameters 
of the lens mass distributions
(such as $\slope$, $\thEin$, $\gext$), $\esrVec$ is a vector of
source-plane pixel intensities, $N_{\rm AGN} (=4)$ is the number of AGN images,
and $\eagnVec_i(\agnVec)$ is the vector of image pixel intensities for
PSF-convolved image $i$ of the AGN.

The likelihood of the ACS data with $N_{\rm d}$ image pixels is 
\bea
\label{eq:like:acs}
\lefteqn{P(\dacsVec | \slope, \thEin, \gext, \parLensVec)} \nonumber
\\
& & = \int
{\rm d}\esrVec\, P(\dacsVec | \slope, \thEin, \gext, \parLensVec,
\esrVec) P(\esrVec),
\eea
where
\bea
\lefteqn{ P(\dacsVec | \slope, \thEin, \gext, \parLensVec, \esrVec) = } \nonumber \\
& & \frac{1}{Z_{\rm d}}  \exp \sum_{i=1}^{N_{\rm d}} \left[
   -\frac{\left(\dacsi - \dacsiP\right)^2}
         {2\sigma_{{\rm pix},i}^2}\right] \nonumber \\
& & \cdot \prod_{i}^{N_{\rm AGN}} \frac{1}{\sqrt{2\pi}\sigma_i}
           \exp\left[-\frac{|\imVec_i-\imVec^{\rm P}_i|^2}{2\sigma_i^2}\right].
\eea
In the first term of this likelihood function,
$Z_{\rm d}$ is just the normalization 
\be
Z_{\rm d} = (2\pi)^{N_{\rm d}/2} \prod_{i=1}^{N_{\rm d}} \sigma_{{\rm
    pix},i},
\ee
$\dacsi$ is the surface brightness of pixel $i$, 
$\dacsiP(\slope, \thEin, \gext, \parLensVec, \esrVec)$ is the corresponding predicted value given by 
\eref{eq:dataP} (recall that $\parLensVec$ are
the remaining lens model parameters to which the ACS 
data are sensitive), and $\sigma_{{\rm pix},i}^2$ 
is the pixel uncertainty given by \eref{eq:noiseboost}.
The second term in the likelihood accounts for the 
positions of the AGN images, modeled as independent 
points (i.e.\ non-pixelated sources) in the image.  
In this term, $\imVec_i$ is the measured image position (listed in
Table \ref{tab:lenspar}), $\sigma_i$ is the
estimated positional uncertainty of $0.005''$, and 
$\imVec^{\rm P}_i(\slope, \thEin, \gext, \parsVec)$ is
the predicted image position given the lens parameters.
(Notice that this second term does not contribute to the
marginalization integral of \eref{eq:like:acs}.)
The form
of $P(\esrVec)$ for the source intensity pixels and the resulting analytic
expression for the marginalization in \eref{eq:like:acs} are
detailed in \citet{SuyuEtal06}.

\begin{table}[!t]
\caption{Lens model parameters}
\label{tab:lenspar}
\begin{center}
\begin{tabular}{l l l}
\hline
& & Marginalized \\
Description & Parameter$\phantom{00}$ &  or optimized \\
& & constraints \\
\hline
\hline
Time-delay distance (Mpc) & $\tdistmod$ & $1883^{+89}_{-85}$\\
\hline
\hline
Lens mass distribution & & \\
\hline
Centroid of G in $\theta_1$ (arcsec) & $\theta_{\rm 1,G}$ & $4.420^{+0.003}_{-0.002}$ \\
Centroid of G in $\theta_2$ (arcsec) & $\theta_{\rm 2,G}$ & $3.932^{+0.004}_{-0.003}$ \\
Axis ratio of G & $q_{\rm G}$ & $0.763 ^{+0.005}_{-0.008}$\\
Position angle of G ($\degr$) & $\phi_{\rm G}$ & $115.8^{+0.5}_{-0.5}$\\
Einstein radius of G (arcsec) & $\thEin$ & $1.64^{+0.01}_{-0.02}$\\
Radial slope of G & $\slope$ & $1.95^{+0.05}_{-0.04}$\\
Centroid of S in $\theta_1$ (arcsec) & $\theta_{\rm 1,S}$ & $4.323$ \\
Centroid of S in $\theta_2$ (arcsec) & $\theta_{\rm 2,S}$ & $4.546$ \\
Einstein radius of S (arcsec) & $\thEinS$ & $0.20^{+0.01}_{-0.01}$\\
External shear strength & $\gext$ & $0.089^{+0.006}_{-0.006}$\\
External shear angle ($\degr$) & $\phi_{\rm ext}$ & $92^{+1}_{-2}$ \\
\hline
\hline
Lens light as S{\'e}rsic profiles & & \\
\hline
Centroid of G in $\theta_1$ (arcsec) & $\theta_{\rm 1,GL}$ & $4.411^{+0.001}_{-0.001}$ \\
Centroid of G in $\theta_2$ (arcsec) & $\theta_{\rm 2,GL}$ & $4.011^{+0.001}_{+0.001}$\\
Position angle of G ($\degr$) & $\phi_{\rm GL}$ & $121.6^{+0.5}_{-0.5}$\\
Axis ratio of G1  & $q_{\rm GL1}$ & $0.878^{+0.004}_{-0.003}$\\
Amplitude of G1 & $A_{\rm GL1}$ & $0.091^{+0.001}_{-0.001}$ \\
Effective radius of G1 (arcsec) & $R_{\rm eff,GL1}$ & $2.49^{+0.01}_{-0.01}$ \\
Index of G1 & $n_{\rm sersic,GL1}$ & $0.93^{+0.03}_{-0.03}$\\
Axis ratio of G2 & $q_{\rm GL2}$ & $0.849^{+0.004}_{-0.004}$\\
Amplitude of G2 & $A_{\rm GL2}$ & $0.89^{+0.03}_{-0.03}$ \\
Effective radius of G2 (arcsec) & $R_{\rm eff,GL2}$ & $0.362^{+0.009}_{-0.009}$ \\
Index of G2 & $n_{\rm sersic,GL2}$ & $1.59^{+0.03}_{-0.03}$ \\
Centroid of S in $\theta_1$ (arcsec) & $\theta_{\rm 1,SL}$ & $4.323$  \\
Centroid of S in $\theta_2$ (arcsec) & $\theta_{\rm 2,SL}$ & $4.546$ \\
Axis ratio of S & $q_{\rm SL}$ & $\equiv 1$\\
Amplitude of S & $A_{\rm SL}$ & $34.11$ \\
Effective radius of S (arcsec) & $R_{\rm eff,SL}$ & $0.01$ \\
Index of S & $n_{\rm sersic,SL}$ & $\equiv 1$ \\
\hline
\hline
Lensed AGN light & & \\
\hline
Position of image A in $\theta_1$ (arcsec) & $\theta_{\rm 1,A}$ & $2.383$ \\
Position of image A in $\theta_2$ (arcsec) & $\theta_{\rm 2,A}$ & $3.412$ \\
Amplitude of image A & $a_{\rm A}$ & $1466$ \\
Position of image B in $\theta_1$ (arcsec) & $\theta_{\rm 1,B}$ & $2.344$\\
Position of image B in $\theta_2$ (arcsec) & $\theta_{\rm 2,B}$ & $4.594$\\
Amplitude of image B & $a_{\rm B}$ & $1220$\\
Position of image C in $\theta_1$ (arcsec) & $\theta_{\rm 1,C}$ & $2.960$\\
Position of image C in $\theta_2$ (arcsec) & $\theta_{\rm 2,C}$ & $2.300$\\
Amplitude of image C & $a_{\rm C}$ & $502$\\
Position of image D in $\theta_1$ (arcsec) & $\theta_{\rm 1,D}$ & $5.494$\\
Position of image D in $\theta_2$ (arcsec) & $\theta_{\rm 2,D}$ & $4.288$\\
Amplitude of image D & $a_{\rm D}$ & $129$\\
\hline
\end{tabular}
\end{center}
Notes. There are a total of 39 parameters that are optimized or
sampled.  Parameters that are optimized are held fixed in the sampling
of the full parameter space and have no uncertainties tabulated.
Changes in these optimized parameters have little
effect on the key parameters for cosmology (such as $\tdistmod$).  The
tabulated values for the sampled parameters are the marginalized
constraints with uncertainties given by
the $16^{\rm th}$ and $84^{\rm th}$ percentiles (to indicate the 68\%
credible interval).  For the
lens light, two S{\'e}rsic profiles with common centroid and position
angle are used to describe the primary lens galaxy G.  They are
denoted as G1 and G2 above. 
The position angles are measured counterclockwise from positive
$\theta_2$ (north). 
The source surface brightness of the AGN host is modeled on a grid of
pixels; these pixel parameters ($\esrVec$) are analytically
marginalized and are thus not listed.
\end{table}

The likelihood for the time delays is given by 
\bea
\label{eq:like:dt}
\lefteqn{ P(\dtVec | \tdistmod, \slope, \thEin, \gext, \parLensVec) = } \nonumber \\
& & \prod_{i} \left( \frac{1}{\sqrt{2\pi}\sigma_{\dt,i}}
                     \exp \left[ \frac{(\dt_i-\dt_i^{\rm P})^2}
                                      {2\sigma_{\dt,i}^2} \right] \right),
\eea
where $\dt_i$ is the measured time delay with uncertainty
$\sigma_{\dt,i}$ for image pair $i$=AB, CB, or DB, and
$\dt_i^{\rm P}(\tdistmod,\slope,\thEin,\gext,\parLensVec )$ is the 
corresponding predicted time delay computed via 
\eref{eq:dt} given the lens mass model parameters.  

The joint likelihood for the ACS and time delay data that appears in
\eref{eq:postParAll}, $P(\dacsVec, \dtVec | \tdist,
\slope, \thEin, \gext, \parLensVec, \kext)$, is just the product of the
likelihoods in Equations (\ref{eq:like:acs}) and (\ref{eq:like:dt}).

We assign uniform priors over reasonable linear ranges for 
all the lens parameters: $\tdistmod$,
$\slope$, $\thEin$, $\gext$ and $\parLensVec$.  In particular, for the
first four lens parameters, the linear ranges for the priors are $\tdistmod
\in [0,10000]\, {\rm Mpc}$, $\slope \in [1.5,2.5]$, $\thEin \in
[0,5]''$, and $\gext \in [0,1]$.


\subsection{MCMC sampling}
\label{sec:lensmod:mcmc}

\begin{figure*}[t]
  \centering
  \includegraphics[width=0.65\textwidth]{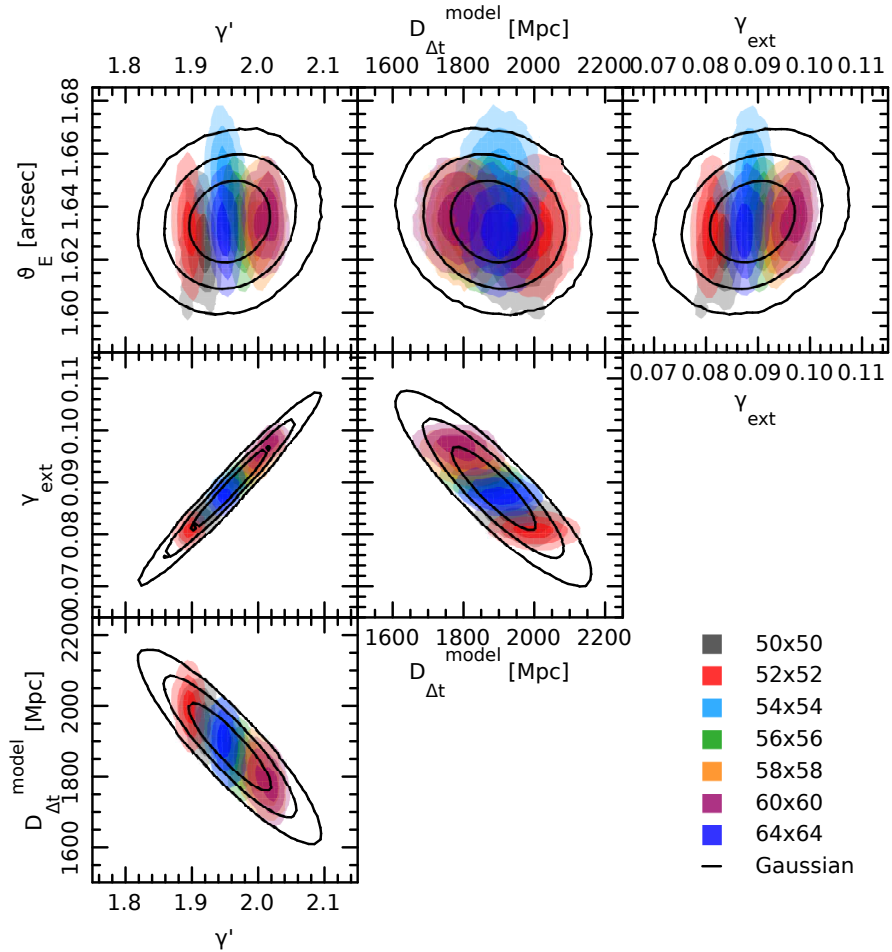}
  \caption{\label{fig:lenschains} Posterior of the key lens model
    parameters for the lensing and time-delay data.  Each color
    represents a particular source resolution that is the dominant
    systematic uncertainty in the modeling of the ACS data.
    The solid curves are a Gaussian fit to the PDF by weighting each
    source resolution chain equally.  The contours/shades mark the
    68.3\%, 95.4\%, 99.7\% credible regions.
    }
\end{figure*}

We model the ACS image and time delays with
{\sc Glee}, a software package developed by 
  \citet{SuyuHalkola10} based on \citet{SuyuEtal06} and \citet{HalkolaEtal08}.
The ACS image has $160$$\times$$160$
surface brightness pixels as constraints.  There are a total of 39 lens
model parameters that are summarized in \tref{tab:lenspar}.  This
is the most comprehensive lens model of \rxj\ to
date.  The list of parameters excludes
the source surface brightness pixel parameters, $\esrVec$, which are
analytically marginalized in computing the likelihood \citep[see,
e.g.,][for details]{SuyuHalkola10}.  With such a large parameter
space, we sequentially sample individual parts of the parameter space
first to get good starting positions near the peak of the PDF before
sampling the full parameter space.  The aim is to obtain a robust PDF
for the key lens parameters for cosmography: $\tdistmod$, $\slope$,
$\thEin$, and $\gext$.

For an initial model of the lens light, we create an annular mask for
the lensed arc and use the image pixels outside the annular mask to
optimize the lens S{\'e}rsic profiles.  The parameters for the light
of the satellite is fixed to these optimized values for the remainder
of the analysis since the satellite light has negligible effect on
$\tdistmod$ and other lens parameters.  Furthermore, we fix the
centroid of the satellite's mass distribution to its observed light
centroid.  We obtain an initial mass model for the lenses using the
image positions of the multiple knots in the source that are
identified following \citet{BrewerLewis08}.  Specifically, we optimize
for the parameters that minimize the separation between the
identified image positions and the predicted image positions from the
mass model.  We then optimize the AGN light together with the light of
the extended source while keeping the lens light and lens mass model
fixed.  The AGN light parameters are then held fixed to these
optimized values.  Having obtained initial values for all the lens
model parameters to describe the ACS data, we then proceed to sample
the lens parameters listed in \tref{tab:lenspar} using a MCMC method.
In particular, we simultaneously vary the
following parameters: modeled time-delay distance, all mass parameters of G, the Einstein radius of
S, external shear, the extended source intensity distribution, and the
lens light profile of G.  {\sc Glee} employs several of the methods of
\citet{DunkleyEtal05} for
efficient MCMC sampling and for assessing chain convergence.


\subsection{Constraints on the lens model parameters}
\label{sec:lensmod:constraints}

\begin{figure*}
  \centering
  \includegraphics[width=0.90\textwidth]{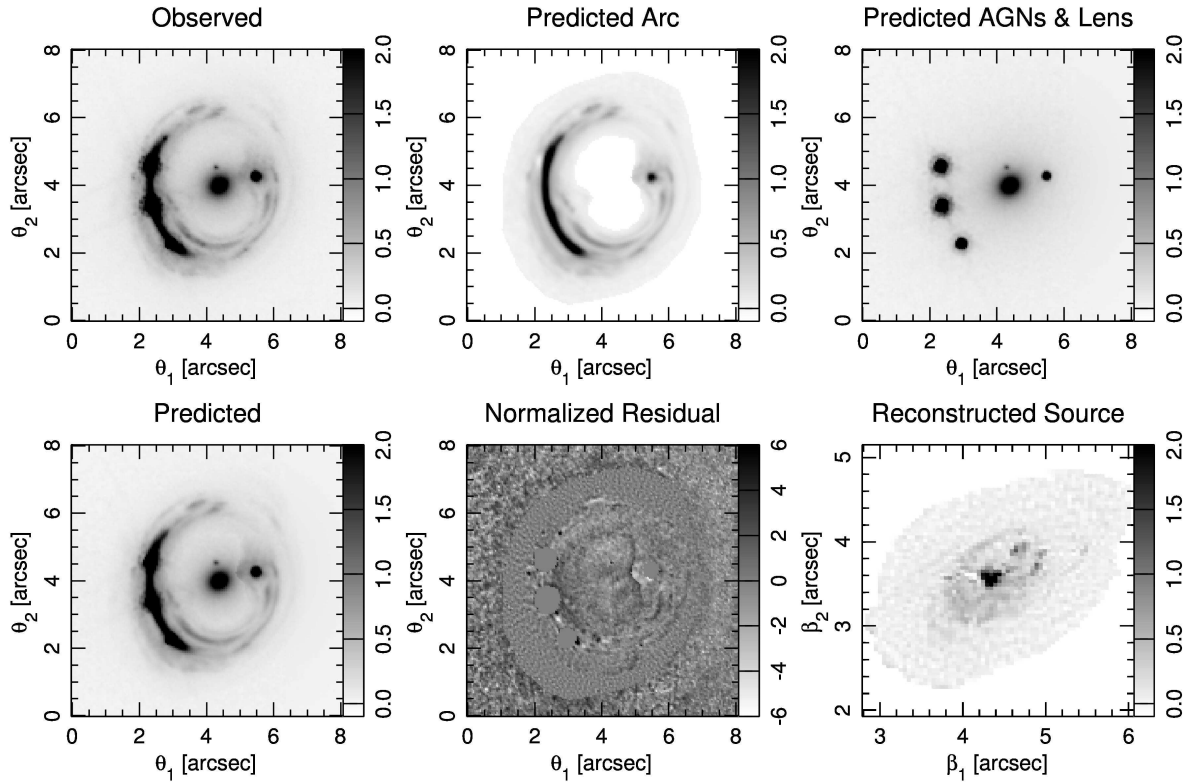}
  \caption{\label{fig:acsrecon} ACS image reconstruction of the most
    probable model with a source grid of 64$\times$64 pixels.  Top left:
    observed ACS F814W image.  Top middle: predicted lensed image
    of the background AGN host galaxy.  Top right: predicted light
    of the lensed AGNs and the lens galaxies.  Bottom left: predicted
    image from all components, which is a sum of the top-middle and
    top-right panels.  Bottom middle: image residual, normalized by
    the estimated 1$\sigma$ uncertainty of each pixel.  Bottom right:
    the reconstructed host galaxy of the AGN in the source
      plane.  Our lens model reproduces the global features of the
    data.}
\end{figure*}

We explore various parameter values for the AGN light and the
satellite S{\'e}rsic light, try different PSF models, and consider
different masks for the lensed arcs and the AGNs (which are fixed in
the MCMC sampling).  These variations
have negligible effect on the sampling of the other lens parameters.  The only 
attribute that changes the PDF of the parameters significantly is the
number of source pixels, or equivalently, the source pixel size.  
We try a series of source resolution from coarse to fine, and the
parameter constraints stabilize starting at $\sim$50$\times$50 source pixels,
corresponding to source pixel sizes of $\sim$$0\farcs05$.  Nonetheless, 
the parameter constraints for different source resolutions are shifted
significantly from one another.  Different source pixelizations
minimize the image residuals in different manners, and predict
different relative thickness of the arcs that provides information on
the lens profile slope $\slope$ \citep[e.g.,][]{Suyu12}.  To quantify
this systematic uncertainty, we consider the 
following set of source resolutions: $50$$\times$$50$, $52$$\times$$52$,
$54$$\times$$54$, $56$$\times$$56$, $58$$\times$$58$, $60$$\times$$60$, and $64$$\times$$64$.
The likelihood $P(\dacsVec, \dtVec | \tdistmod, \slope, \thEin,
\gext)$, which is proportional to the marginalized posterior of these
parameters $P(\tdistmod, \slope, \thEin, \gext | \dacsVec, \dtVec)$
since the priors are uniform, is plotted in \fref{fig:lenschains}
for each of these source resolutions.  The scatter in constraints among
the various source resolutions allows us to quantify the systematic
uncertainty.  In particular, we weight each choice of the source
resolution equally, and combine the Markov chains together.  In 
\tref{tab:lenspar}, we list the marginalized parameters from the
combined samples.

We show the most probable image and source reconstruction for the
$64$$\times$$64$ resolution in \fref{fig:acsrecon}.  Only the image
intensity pixels within the annular mask shown in the top-middle panel
are used to reconstruct the source that is shown in the bottom-right
panel.  A comparison of the top-left and bottom-left panels shows that
our lens model reproduces the global features of the ACS image.
The time delays are also reproduced by the model: for the various
source resolutions, the $\chi^2$ (not reduced) is $\sim$2
for the three delays relative to image B. There
are some small residual features in the bottom-middle panel of
\fref{fig:acsrecon}, and these 
cause the shifts in the parameter constraints seen in
\fref{fig:lenschains} for different source pixel sizes.  The
reconstructed host galaxy of the AGN in the bottom-right panel shows a
compact central peak, which is probably the bulge of the spiral source
galaxy, embedded in a more diffuse patch of light (the disk) with 
knots/spiral features.  The bulge and disk have half-light radii of
$\sim$$0\farcs1$ and $\sim$$0\farcs8$, respectively.  Given the source
redshift, this implies a bulge size of $\sim$0.7\,kpc and a disk size of
$\sim$5\,kpc, which are typical for disk galaxies at these redshifts
\citep[e.g.,][]{BardenEtal05, MacArthurEtal08} and are comparable to
the largest sources in the lens systems of the Sloan Lens ACS survey
\citep{NewtonEtal11}.


\subsection{Understanding the external shear}
\label{sec:lensmod:shear}

The inferred external shear is $\gext=0.089\pm0.006$ (marginalizing over
all other model parameters) from modeling the ACS image and the time
delays.  The 
external shear may provide information on the amount of external
convergence, since they originate from the same external structures.
However, the high $\gext$ found in our model could potentially be
attributed to deviations of  the primary lens from its elliptical power-law
description; if this were the case, some of $\gext$ would in fact be internal
shear.  To gauge whether the modeled shear is
truly external, we also considered a model that includes a constant
external convergence gradient.  This introduces two additional
parameters: $\kappa'$ (gradient) and $\phi_{\kappa}$ (the position
angle of the gradient, where $\phi_{\kappa}=0$ corresponds to positive
$\kappa$ gradient along the positive $\im_2$ direction, i.e., north).  The ACS
data allow us to constrain $\kappa'=(5.1^{+0.4}_{-0.3})\times10^{-3}\,
\rm{arcsec}^{-1}$ and $\phi_{\kappa}=87\pm2\degr$.  The convergence
gradient is aligned along the same direction 
as the external shear within $5\degr$ and has a sensible magnitude,
suggesting that the shear is in fact
truly external, and is likely due to 
mass structures to the east of the lens.

\begin{figure*}
  \centering
  \includegraphics[width=0.90\textwidth]{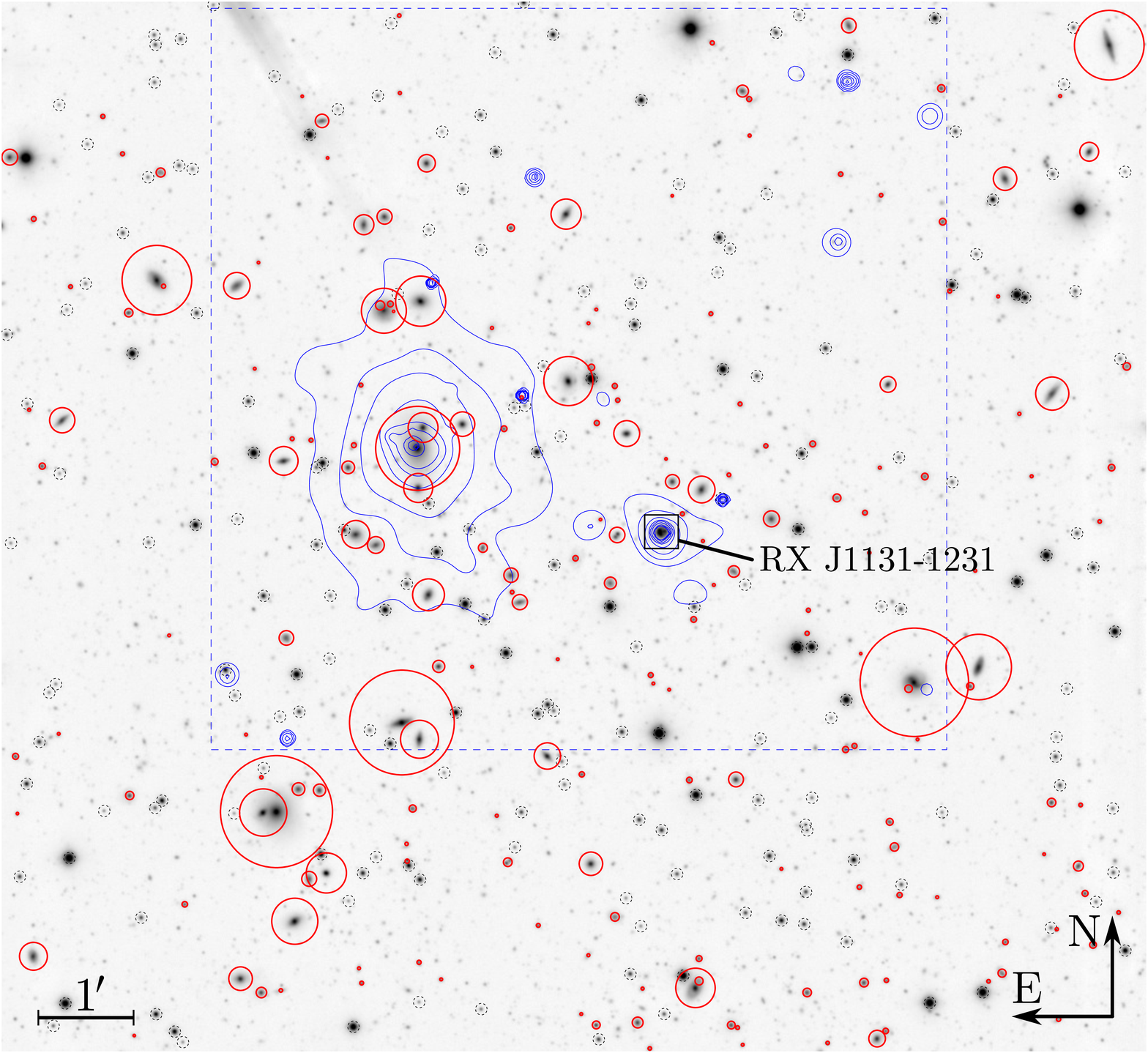}
  \caption{\label{fig:eulerstack} $11.5'$$\times$$10.5'$ R-band image
    obtained from stacking 60 hours of the best-quality images in the
    COSMOGRAIL monitoring.  The lens system is marked by the box near
    the center.  Galaxies (stars) in the field are indicated by solid
    (dashed) circles.  The radius of the solid circle is proportional
    to the flux of the galaxy.  X-ray map from \citet{ChartasEtal09}
    are overlaid on the image within the dashed box.  The
    concentrations of mass structures to the east of the lens are
    consistent with the modeled external shear and convergence
    gradient directions. }
\end{figure*}

To investigate the origin of the external shear, 
we construct a wide-field R-band image from the COSMOGRAIL monitoring
images that is shown in \fref{fig:eulerstack}.  The lens system is
indicated by the box, and the galaxies (stars) in the field are marked
by solid (dashed) circles, identified using SExtractor
\citep{BertinArnouts96}.  Overlaid on the image within the dashed box
are the X-ray contours from \citet{ChartasEtal09}, showing the
presence of a galaxy cluster 
that is located at $158''$
northeast of the lens \citep{MorganEtal06, ChartasEtal09}.  
The cluster is at $z=0.1$ based on the redshift measurements
  of two of the red-sequence cluster galaxies from the Las Campanas
  Redshift Survey \citep{ShectmanEtal96, WilliamsEtal06}.  Using the
  measured 2$-$10\,keV luminosity of $1.7\times 10^{42} {\rm
    \,ergs\,s^{-1}}$, X-ray temperature of 1.2\,keV and core radius of
  a $\beta$ model of $4.2''$ for the cluster \citep{ChartasEtal09}, we
  estimate that the contribution of the cluster to the external shear
  at the lens is only a few percent.
Nonetheless, large-scale structures associated with
the cluster and the plethora of mass structures to the east of the
lens could generate additional shear.  The fact that our modeled
external shear and convergence gradients both point toward mass structures
in the east that are visible in \fref{fig:eulerstack} is a further
indication that the modeled shear is indeed external.  We will use this
external
shear in \sref{sec:breakmsd:env} to constrain the
external convergence.


\subsection{Propagating the lens model forward}
\label{sec:lensmod:propagation}

To facilitate the sampling and marginalization of the posterior of the
{\it cosmological} parameters in Equations (\ref{eq:postParAll}) and
(\ref{eq:postCosmo}), we approximate the overall likelihood of $\dacsVec$
and $\dtVec$ from the multiple source resolutions in
\fref{fig:lenschains} with a multivariate Gaussian 
distribution for the interesting parameters $\slope, \thEin, \gext$ and
$\tdistmod$, marginalizing over the nuisance parameters $\parLensVec$. 
This approximation allows the value of 
$P(\dacsVec,\dtVec|\slope,\thEin,\gext,\tdistmod)$ to be computed at any
position in this 4-dimensional parameter space.
Note that in contrast to the other parameters, the
Einstein radius of the primary lens galaxy, $\thEin$, is well determined, with
minimal degeneracy with other parameters.  This robust quantity is
used in the dynamics modeling of the lens galaxy.  The approximated
Gaussian likelihood provides an easy way to combine with the stellar
kinematics and lens environment information for measuring $\tdist$.


\section{Constraining the external convergence $\kext$}
\label{sec:breakmsd}

In this section, 
we fold in additional information on the lens galaxy stellar
kinematics and density environment to constrain the nuisance
parameter $\kext$ (which characterizes the effects of LOS
  structures). 


\subsection{Stellar kinematics}
\label{sec:breakmsd:vd}

We follow \citet{SuyuEtal10} and model the velocity dispersion of
the stars in the primary lens galaxy G, highlighting the main steps.
The three-dimensional mass density distribution of the lens galaxy can
be expressed as 
\be
\rho_{\rm G}(r)=(\kext - 1)
\Sigma_{\rm crit} \thEin^{\slope-1} D_{\rm d}^{\slope-1}\frac{\Gamma(\frac{\slope}{2})}{\pi^{1/2}
  \Gamma(\frac{\slope-3}{2})} \frac{1}{r^{\slope}}.
\ee
Note that the projected mass of the lens galaxy enclosed
within $\thEin$ is $(1-\kext) M_{\rm E}$, while 
the projected mass associated with the
external convergence is $\kext M_{\rm E}$; the sum of the two is the
Einstein mass
$M_{\rm E}$ that was given in \eref{eq:MEin}.  We employ
spherical Jean's modeling to infer the line-of-sight velocity
dispersion, $\vdisp^{\rm P}(\parCosVec, \slope, \thEin, \rani, \kext)$,
from $\rho_{\rm G}$ by assuming the Hernquist profile
\citep{Hernquist90} for the stellar distribution
\citep[\eg][]{BinneyTremaine87,SuyuEtal10}.\footnote{
  \citet{SuyuEtal10}
  found that \citet{Hernquist90} and \citet{Jaffe83} stellar
  distribution functions led to nearly identical cosmological
  constraints.}
An anisotropy radius of $\rani=0$ corresponds to
pure radial stellar orbits, while $\rani\rightarrow\infty$ 
corresponds to isotropic orbits with
equal radial and tangential velocity dispersions.  We note that
$\vdisp^{\rm P}$ is independent of $H_0$, but is dependent on the 
other cosmological parameters (\eg $w$ and $\Ode$) through 
$\Sigma_{\rm crit}$ and the physical scale radius of the stellar 
distribution.

The likelihood for the velocity dispersion is 
\bea
\label{eq:vdisplike}
\lefteqn{ P(\vdisp | \parCosVec, \slope, \thEin, \rani, \kext) } \nonumber \\
& & = \frac{1}{\sqrt{2\pi\sigma_{\vdisp}^2}} \exp{\left[-\frac{(\vdisp -
\vdisp^{\rm P}(\parCosVec, \slope, \thEin, \rani, \kext))^2}{2\sigma_{\vdisp}^2}\right]},
\eea
where $\vdisp=323\,\kms$ and $\sigma_{\vdisp}=20\,\kms$ from
\sref{sec:obs:vd}.  Recall that the priors on $\slope$ and $\thEin$
were assigned to be uniform in the lens modeling.  
We also impose a uniform prior on
$\rani$ in the range of $[0.5,5]R_{\rm eff}$ for the kinematics
modeling, where the effective
radius based on the two-component S{\'e}rsic profiles in 
\tref{tab:lenspar} is $1\farcs85$ 
from the photometry.\footnote{Before unblinding, we used
an effective radius of $3\farcs2$ based on a single S{\'e}rsic fit.
The larger $R_{\rm eff}$ changes the inference of $\tdist$ at the
$<0.5\%$ level.}
The uncertainty in $R_{\rm eff}$ has negligible
impact on the predicted velocity dispersion.  The prior PDF for
$\parCosVec$ is discussed in \sref{sec:tdist:priors}, while the PDF
for $\kext$ is described in the next section.


\subsection{Lens environment}
\label{sec:breakmsd:env}

We combine the relative galaxy counts from 
\sref{sec:obs:env}, the measured external shear in 
\sref{sec:lensmod:constraints}, and the Millennium Simulation
\citep[MS;][]{SpringelEtal05} to obtain an estimate of
$P(\kext | \denvVec, \gext,{\rm MS})$.  
This builds on the approach presented in
\citet{SuyuEtal10} that used only the relative galaxy counts. 

Tracing rays through the Millennium Simulation \citep[see][for
details of the method]{HilbertEtal09}, we create 64 simulated survey 
fields, each of solid angle $4$$\times$$4\,\degt^2$. 
In each field we map the
convergence and shear to the source redshift $\zs$, and catalog 
the galaxy content, which we derive from the galaxy model by
\citet{GuoEtal2010}. For each line of sight in each simulated
field, we record the convergence, shear, and relative galaxy counts in
a $45''$ aperture having I-band magnitudes between 18.5 and 24.5.
These provide 
samples for the PDF $P(\kext, \gext, \denvVec | {\rm
 MS})$. We assume that the constructed PDF is applicable to
strong-lens lines of sight, following \citet{SuyuEtal10} who
showed that the
distribution of $\kext$ for strong lens lines of sight are very
similar to that for all lines of sight.  

Structures in front of the lens distort the
time delays and the images of the lens/source, 
while structures behind the
lens further affect the time delays and images of the source. 
However, to model simultaneously the mass distributions of the
strong lens galaxies and all structures along the line of sight 
is well beyond current capabilities.
In practice, the modeling of the strong lens galaxies is performed
separately from the description of line-of-sight structures, and we
approximate the effects of the lines-of-sight structures into the 
single correction term $\kext$, whose statistical properties we
estimate from the Millennium Simulation. 

\begin{figure}
  \centering
  \includegraphics[width=0.45\textwidth, clip]{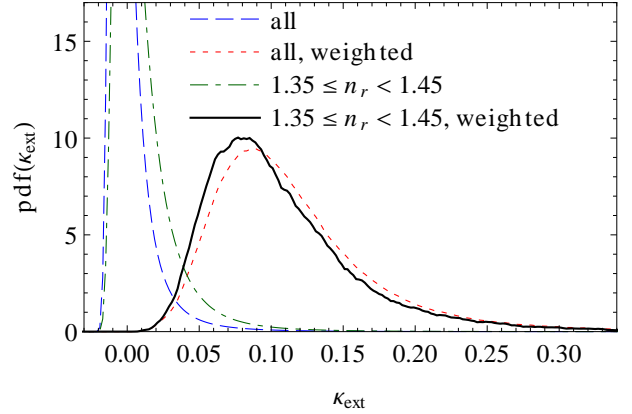}
  \caption{\label{fig:kextpdf} The effective prior probability distribution 
    for the external convergence $\kext$ from combining ray tracing
    through the Millennium Simulation with (1) the galaxy
    count around the lens system relative to the average number of
    galaxy counts, and/or (2) the modeled external shear of $0.089\pm0.006$.
    Dashed line: the convergence distribution for all lines of sight;
    Dot-dashed line: the convergence distribution for lines of sight with relative galaxy count $n_{\rm r} = 1.4\pm0.05$;
    Dotted line: the convergence distribution for all lines of sight weighted by the likelihood for $\gext$ from the lens model;
    Solid line: the $\gext$-weighted convergence distribution for lines of sight with $n_{\rm r}=1.4\pm0.05$. 
    The effective prior for $\kext$ used in the final cosmological
    parameter inference is described 
    by this, most informative, distribution.}
\end{figure}

By selecting the lines of sight in the Millennium Simulation that
match the
properties of \rxj, we can obtain $P(\denvVec | \kext,\gext,{\rm
  MS})\,P(\kext)$ and simultaneously marginalize over $\gext$ in
\eref{eq:postParAll}.  We assumed a uniform prior for
$\gext$ in the lensing analysis, such that $P(\gext)$ is a constant.  The
lensing likelihood is the only other term that depends on $\gext$, and
from \sref{sec:lensmod:constraints}, the lensing likelihood 
provides a tight constraint on $\gext$ that is approximately Gaussian:
$0.089\pm0.006$.  We can therefore simplify part of 
\eref{eq:postParAll} to 
\bea
\lefteqn \nonumber & \int {\rm d}\gext P(\dacsVec, \dtVec
| \tdist, \slope, \thEin, \gext, \kext) \\
\nonumber &  \ \ \ \ \cdot P(\denvVec | \kext, \gext, {\rm MS}) \\
\lefteqn \nonumber & \simeq P(\dacsVec, \dtVec | \tdist, \slope, \thEin,
\kext)\\
& \ \ \ \ \cdot P(\denvVec | \kext, \gext=0.089\pm0.006, {\rm MS}),
\eea
where the above approximation, i.e., neglecting the covariance between $\gext$ 
and the other parameters in the lensing likelihood and then marginalizing
$\gext$ separately, is conservative since we would gain in
precision by including the covariances with other parameters.
Furthermore, by Bayes' rule,
\bea
\lefteqn \nonumber P(\denvVec | \kext, \gext=0.089\pm0.006, {\rm MS})
P(\kext) \\
\propto P(\kext | \denvVec, \gext=0.089\pm0.006, {\rm MS}),
\eea
which is precisely the PDF of $\kext$ by selecting the samples in 
$P(\kext, \gext, \denvVec | {\rm MS})$ that satisfies $\denvVec$ with
a relative galaxy count within $1.4\pm0.05$, and subsequently weighting
these samples by the Gaussian likelihood for $\gext$.  This effective
prior PDF for $\kext$ that is constructed from
the weighted samples, $P(\kext 
| \denvVec, \gext=0.089\pm0.006, {\rm MS})$, is shown by the solid line in Figure
\ref{fig:kextpdf}.

\section{Time-delay distance of \rxj}
\label{sec:tdist}

We combine all the PDFs obtained 
in the previous sections to infer the
time-delay distance $\tdist$.


\subsection{Cosmological priors}
\label{sec:tdist:priors}

As written above, we {\it could} infer the time delay distance $\tdist$
directly, given a uniform prior. However, we are primarily interested in the
cosmological information contained in such a distance measurement, so prefer
to infer these directly.
The posterior probability distribution on $\tdist$ can then 
be obtained by
first calculating the posterior PDF of the cosmological parameters
$\parCosVec$ through the marginalizations in Equations
(\ref{eq:postCosmo}) and (\ref{eq:postParAll}), and then changing variables to
$\tdist$. Such transformations are of course straightforward when working with
sampled PDFs. 

As described in \tref{tab:cosmoprior}, we consider the following
five cosmological world models, each with its own 
prior PDF $P(\parCosVec)$: U$H_0$, U$w$CDM, WMAP7$w$CDM,
WMAP7o$\Lambda$CDM, and WMAP7o$w$CDM.


\subsection{Posterior sampling}
\label{sec:tdist:comb}

We sample the posterior PDF by weighting samples drawn from the prior
PDF with the joint likelihood function evaluated at those points
\citep{SuyuEtal10}.
We generate samples of the cosmological
parameters $\parCosVec$ from the priors listed in 
\tref{tab:cosmoprior}.  We then join these to samples of $\kext$ 
drawn from $P(\kext)$ from \sref{sec:breakmsd:env} and 
shown in \fref{fig:kextpdf}, and to uniformly
distributed samples of $\slope$ within $[1.5,2.5]$ and $\rani$ within
$[0.5,5]\,R_{\rm eff}$.  Rather than generating samples of
$\thEin$ from the uniform prior, we obtain samples of $\thEin$ directly
from the Gaussian approximation to the lensing and time-delay
likelihood since $\thEin$ is quite independent of other model
parameters (as shown in \fref{fig:lenschains}).  This boosts
sampling efficiency, and the $\thEin$ samples are only used to evaluate
the kinematics likelihood.   

For each sample of \{$\parCosVec$, $\kext$, $\slope$, $\rani$, $\thEin$\}, we
obtain the weight (or importance) as follows: (1) we determine $\tdist$ from
$\parCosVec$ via \eref{eq:tdist}, (2) we calculate $\tdistmod$ via
\eref{eq:msd:tdist}, (3) we evaluate $P(\dacsVec, \dtVec | \tdistmod, \slope)$ based on the Gaussian approximation shown in
\fref{fig:lenschains} for $\tdistmod$ and $\slope$,
(4) we compute $P(\vdisp | \parCosVec, \slope, \kext, \thEin$, $\rani)$ via
\eref{eq:vdisplike}, and (5) we weight the sample by the product of
$P(\dacsVec, \dtVec | \tdistmod, \slope)$ and $P(\vdisp | \parCosVec,
\slope, \kext, \thEin$, $\rani)$ from the previous two steps.  The projection
of these weighted samples onto $\parCosVec$ or $\tdist$ effectively
marginalizes over the other parameters.

\begin{table}
\caption{Priors on cosmological parameters}
\label{tab:cosmoprior}
\begin{center}
\begin{tabular}{ll}
\hline
Prior & Description \\
\hline
U$H_0$ & Flat $\Lambda$CDM with: \\
 & uniform $H_0$ in $[0,150]\,\kmsMpc$, \\
 & $\Om=0.27$, $\OL=0.73$, $w=-1$. \\
 & This is similar to the typical priors that \\
 & were assumed in most of the early lensing \\
 & studies, which sought to constrain $H_0$ at \\
 & fixed cosmology. \vspace{0.2cm} \\

U$w$CDM & Flat $w$CDM with: \\
& uniform $H_0$ in $[0,150]\,\kmsMpc$, \\
& uniform $\Ode=1-\Om$ in $[0,1]$, \\
& uniform $w$ in $[-2.5,0.5]$. \vspace{0.2cm} \\

WMAP7$w$CDM$^{\dagger}$ & WMAP7 for \{$H_0$, $\Ode$, $w$\} in $w$CDM \\
 & with flatness and time-independent $w$. \vspace{0.2cm} \\

WMAP7o$\Lambda$CDM$^{\dagger}$ & WMAP7 for \{$H_0$, $\OL$, $\Om$\} in open \\
&  (or rather, non-flat) $\Lambda$CDM with $w=-1$ \\ 
& and $\Ok=1-\OL-\Om$ as the curvature \\ 
& parameter. \vspace{0.2cm}\\ 

WMAP7o$w$CDM$^{\dagger}$ & WMAP7 for \{$H_0$, $\Ode$, $w$, $\Ok$\} in
open \\ 
& $w$CDM with time-independent $w$ and \\
& curvature parameter $\Ok=1-\Ode-\Om$. \\
\hline
\end{tabular}
\end{center}
$^{\dagger}$The prior PDF for the cosmological parameters are taken to be
the posterior PDF from the WMAP 7-year data set \citep{KomatsuEtal11}.
\end{table}



\subsection{Blind analysis in action}
\label{sec:tdist:blind}

As a brief illustration of our blind analysis approach,
we show in the left panel of \fref{fig:tdistpdf} the blinded
plot of the time-delay distance measurement.  
For all cosmological parameters such
as $\tdist$, $\tdistmod$, $H_0$, $w$, $\Om$, etc., we always plotted
their probability 
distribution with respect to the median during the blind analysis.
Therefore, we could use the shape of the PDFs to check our analysis 
and avoid introducing experimenter bias by blinding the absolute
parameter values.  When we marginalized the parameters during the
blind phase, our analysis code also returned the constraints with
respect to the median.  For example, the blinded time-delay distance
for the WMAP7$w$CDM cosmology would be
$0^{+130}_{-120}$\,Mpc.  We used this
particular cosmology as our fiducial world model 
during the blind analysis. 
In the remainder of the paper, we show
the unblinded results of \rxj.  The comparison with the gravitational
lens \blens\ 
and other cosmological probes was performed after we unblinded the
analysis of \rxj; otherwise, the blind analysis would be spoiled by
such a comparison since the results of these previous studies were
already known.


\subsection{Posterior PDF for $\tdist$}
\label{sec:tdist:pdf}

\begin{figure*}
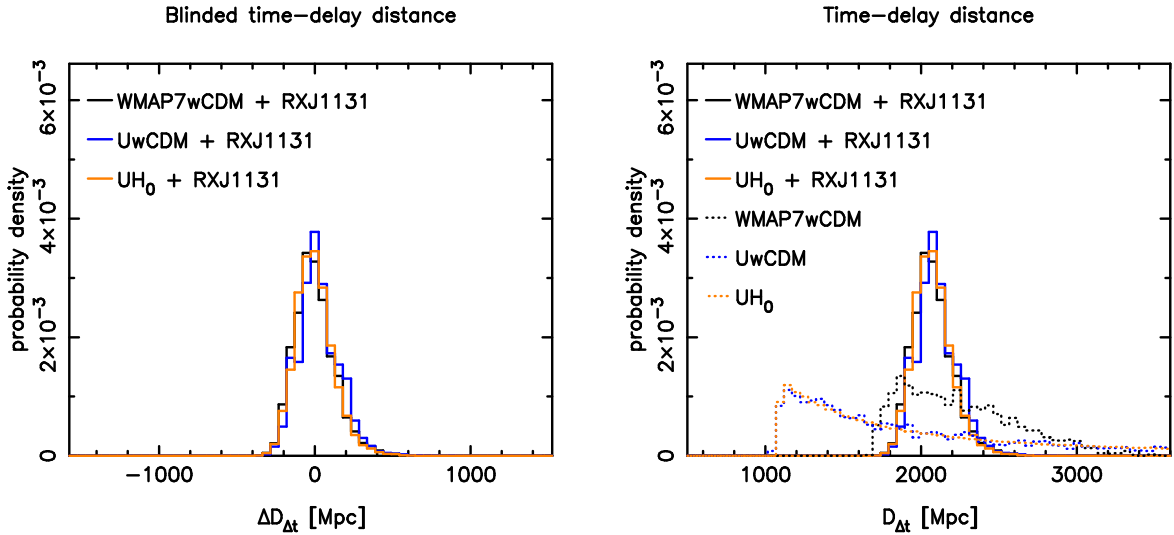

  \centering
  \includegraphics[width=0.45\textwidth, clip]{fig7a.ps}
  \includegraphics[width=0.45\textwidth, clip]{fig7b.ps}
  \caption{\label{fig:tdistpdf} Blinded (left) and unblinded (right)
    PDFs for $\tdist$, showing the \rxj\ posterior constraints on
    $\tdist$ (solid) given assorted priors for the cosmological
    parameters (dotted, labeled).  See 
    \tref{tab:cosmoprior} for a full description of the various
    priors.  \rxj\ provides tight constraints on $\tdist$, which
    translates into information about $\Om$, $\Ode$, $w$ and
    particularly $H_0$.}
\end{figure*}

We show in the right-hand panel of \fref{fig:tdistpdf} the
unblinded probability distribution of the
time-delay distance for the first three cosmological models in
\tref{tab:cosmoprior}.  The priors, shown in dotted lines, are broad
and rather uninformative.  When including information from $\dacsVec$,
$\dtVec$, 
$\vdisp$, and $\kext$, we obtain posterior PDFs of $\tdist$ for \rxj\ that are
nearly independent of the prior, demonstrating that 
time-delay lenses provide
robust measurements of $\tdist$.  We find that the data 
constrain the $\tdist$ to \rxj\ with $\sim$$6\%$ precision.

We can compress these results by approximating the posterior PDF for
$\tdist$ as a shifted log normal distribution:
\bea
\label{eq:DtLogNorm}
\lefteqn{P(\tdist|H_0,\Ode,w,\Om) \simeq }\nonumber\\
 && \frac{1}{\sqrt{2\pi} (x-\lambda_{\rm D}) \sigma_{\rm D}} 
\exp{\left[-\frac{(\log(x-\lambda_{\rm D}) - \mu_{\rm D})^2}{2\sigma_{\rm D}^2}\right]},
\eea
where $x=\tdist/(1\, {\rm Mpc})$, $\lambda_{\rm D} = \Dtfitlam$, $\mu_{\rm
  D}=\Dtfitmu$ and $\sigma_{\rm D} = \Dtfitsig$.  This approximation
accurately reproduces the cosmological inference, in that  $H_0$ is
recovered within $<1\%$ in terms of its median, 16$^{\rm th}$ and
84$^{\rm th}$
percentile values for the WMAP7 cosmologies we have considered.  The
robust constraint on $\tdist$ serves as the basis for cosmological
inferences in Section \ref{sec:cosmo}.

\subsection{Sources of uncertainty}
\label{sec:sourceError}

Our $\tdist$ measurement accounts for all known sources of uncertainty
that we have summarized in \tref{tab:errorbudget}.  The dominant
sources are the first three items.  The precision for the time delay
is the 1$\sigma$ uncertainty as a fraction of the measured value for
the longest delay, $\dt_{\rm DB}$.  For the lens mass model and
line-of-sight contributions, we define the precision as half the
difference between the 16$^{\rm th}$ and 84$^{\rm th}$ percentiles of the PDF for
$\tdistmod$ from Section \ref{sec:lensmod:constraints} in fractions of
its median value and for $\kext$ from Section \ref{sec:breakmsd:env}
in fractions of 1, respectively. 
The remaining sources of uncertainty
are collectively denoted by ``other sources'', and the two main contributors to
this category are the peculiar velocity of the lens and
the impact of the satellite.

Spectroscopic studies of the field of \rxj\ indicate that the lens is
in a galaxy group with a velocity dispersion of $429^{+119}_{-93}\,\kms$
\citep{WongEtal11}.  \rxj\ is the brightest red sequence galaxy in
this group, and is thus likely to be near the center of mass of
the group halo with a small peculiar velocity relative to the
group \citep{ZabludoffMulchaey98, WilliamsEtal06, GeorgeEtal12}.
However, the group could be moving relative to the Hubble flow due to nearby
large-scale structures.  The one-dimensional rms galaxy peculiar
velocity is typically $\lesssim 300 \kms$
\citep[e.g.,][]{MosqueraKochanek11, Peebles93}.  
A peculiar velocity of $300\,\kms$ for \rxj\ would cause
$\tdist$ to change by $0.8\%$.\footnote{The change in redshift due to
  peculiar velocities is described in, e.g., \citet{Harrison74}.}
A similar peculiar velocity for the lensed source has a much smaller impact
on $\tdist$, changing it by only $0.2\%$.   We
note that the peculiar velocities of lenses are stochastic, and
this source of uncertainty should average out in a sample of lenses.

We have explicitly included the satellite in our lens mass model in
\sref{sec:lensmod}.  However, there is some degeneracy in
apportioning the mass between the satellite and the primary lens
galaxy since lensing is mostly sensitive to the total mass enclosed
within the lensing critical curves (approximately traced by the arcs).
The more massive the satellite, the less massive the primary lens
galaxy. Owing to its central location and the degeneracy with the mass
of the main deflector, we expect the impact of the satellite on the
difference in gravitational potential between the multiple images to
be very small.

To assess the effect of the mass of the satellite on our $\tdist$
inference, we consider an extreme model where the satellite has zero
mass.  In this case, we require a more massive primary lens galaxy
with higher $\thEin$ to fit the lensing features, as expected.  The
resulting $\tdistmod$ and $\slope$ from this model are consistent with
that of the original model, but with larger parameter uncertainties
due to poorer fits without the satellite.  Even if we use the
overestimated $\thEin$ of the primary lens from this extreme model for
the kinematics, we find that the effect on the inferred $\tdist$ is at
the $<1\%$ level.

In \tref{tab:errorbudget}, we list the total uncertainty of $6.7\%$
based on a simple Gaussian approximation where we 
add up the uncertainties of each contribution in quadrature.  This is close to the more accurate $6.0\%$ based on
proper sampling that takes into account the non-Gaussian distribution
(e.g., of $\kext$) and the inclusion of the stellar velocity
dispersion.  We note that the sampling does not include explicitly the
other sources that contribute at the $<$$1\%$ level; however, they are
practically insignificant in the overall error budget.  Most of the
uncertainty in $\tdist$ comes from the lens mass model and the
line-of-sight contribution.  Reducing the uncertainty on \rxj's
$\tdist$ would require a better model of the source intensity
distribution that depends less sensitively on the source pixel size
(possibly via an adaptive source pixelization scheme
\citep[e.g.,][]{VegettiKoopmans09}), and a better characterization of
$\kext$ by using more observational information from the
field.  Investigations are in progress to improve $\kext$ constraints
(Greene et al., submitted; Collett et al., submitted).

\begin{table}
\caption{Error budget on time-delay distance of \rxj}
\label{tab:errorbudget}
\begin{center}
\begin{tabular}{lc}
\hline
Description & uncertainty  \\
\hline
time delays & 1.6\% \\
lens mass model & 4.6\% \\
line-of-sight contribution & 4.6\% \\
other sources & $<$1\% \\
\hline
Total (Gaussian approximation) & 6.7\% \\
Total (full sampling) & 6.0\% \\
\hline
\end{tabular}
\end{center}
Notes.  The other sources of uncertainty that contribute at the $<$$1\%$
include the peculiar velocity of the lens and the impact of the
satellite.  Details are in \sref{sec:sourceError}.  The
Gaussian approximation simply adds the uncertainties in quadrature,
providing a crude estimate for the total uncertainty based on the full
sampling of the non-Gaussian PDFs.
\end{table}


\section{Cosmological inference}
\label{sec:cosmo}

We now present our inference on the parameters of the expanding Universe
and compare our results to other cosmographic probes.  Specifically,
our $\tdist$ measurement for \rxj\ provides information on cosmology
that is illustrated in \sref{sec:cosmo:rxj}.  We compare the
results to that of \blens\ to check for consistency in \sref{sec:cosmo:compblens}, before combining the two lenses together in
\sref{sec:cosmo:combinelenses}.  We then compare the constraints
from the two time-delay lenses to a few other cosmological probes in
\sref{sec:cosmo:compprobes}. 


\subsection{Constraints from \rxj}
\label{sec:cosmo:rxj}

We have seen that the \rxj\ $\tdist$ measurement is nearly independent of
assumptions about the background cosmology.  While $\tdist$ is
primarily sensitive to $H_0$, information from $\tdist$ must be shared
with other cosmological parameters via the combination of angular
diameter distances.
Therefore, cosmological parameter constraints will depend somewhat on our
assumptions for the background cosmology.  
In this section we consider the first three
cosmologies listed in \tref{tab:cosmoprior}: U$H_0$, U$w$CDM, and
WMAP7$w$CDM.  

With all other parameters fixed in the U$H_0$ cosmology except for
$H_0$, all our knowledge of $\tdist$ is converted to information on
$H_0$.  We therefore obtain a precise measurement of $H_0= 78.7
^{+4.3}_{-4.5}\,\kmsMpc$ for \rxj\ with a 5.5\% uncertainty.

\begin{figure}[t!]
  \centering
  \includegraphics[width=0.45\textwidth, clip]{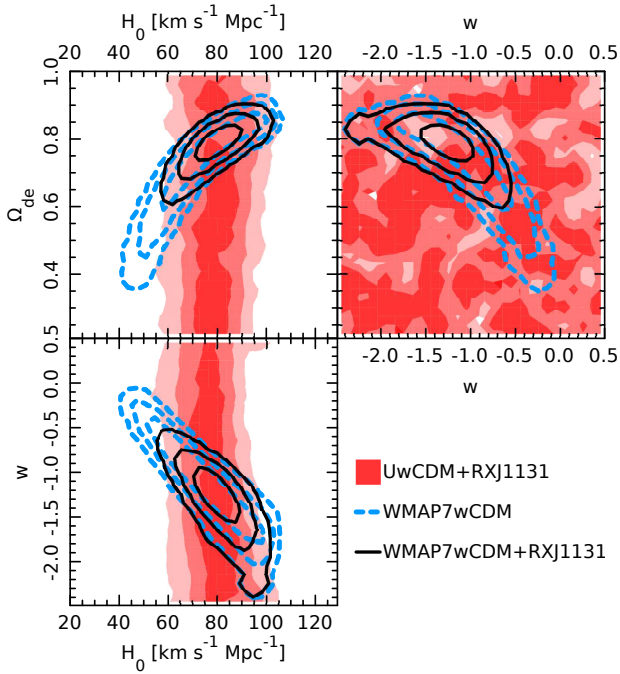}
  \caption{\label{fig:wCDM1131} \rxj\ marginalized posterior PDF for
    $H_0$, $\Ode$ and $w$ in flat $w$CDM cosmological models.
    Contours/shades mark the 68.3\%, 95.4\%, 99.7\% credible
    regions. The three sets of contours/shades correspond to three
    different prior/data set combinations. Shaded red: \rxj\ constraints
    given by the U$w$CDM prior; dashed blue: the prior provided by the WMAP7
    data set alone; solid black: the joint constraints from
    combining WMAP7 and \rxj.}
\end{figure}

Next, we relax our assumptions on $\Ode$, $\Om$ and $w$, and consider
the U$w$CDM and WMAP7$w$CDM cosmologies in a flat Universe.
\fref{fig:wCDM1131} shows the resulting constraints.
The contours for the U$w$CDM cosmology with vertical 
bands in the $H_0$ panels illustrates that the time-delay distance is
mostly sensitive to $H_0$.  The constraint on $H_0$ breaks the parameter
degeneracies in the WMAP7 data set, and we obtain the following joint
parameter constraints for \rxj\ in combination with WMAP7:
$H_0=80.0^{+5.8}_{-5.7}\,\kmsMpc$, $\Ode=0.79\pm0.03$, and
$w=-1.25^{+0.17}_{-0.21}$.  

\subsection{Comparison between \rxj\ and \blens}
\label{sec:cosmo:compblens}

\begin{figure}
  \centering
  \includegraphics[width=0.45\textwidth, clip]{fig9a.ps}
  \includegraphics[width=0.45\textwidth, clip]{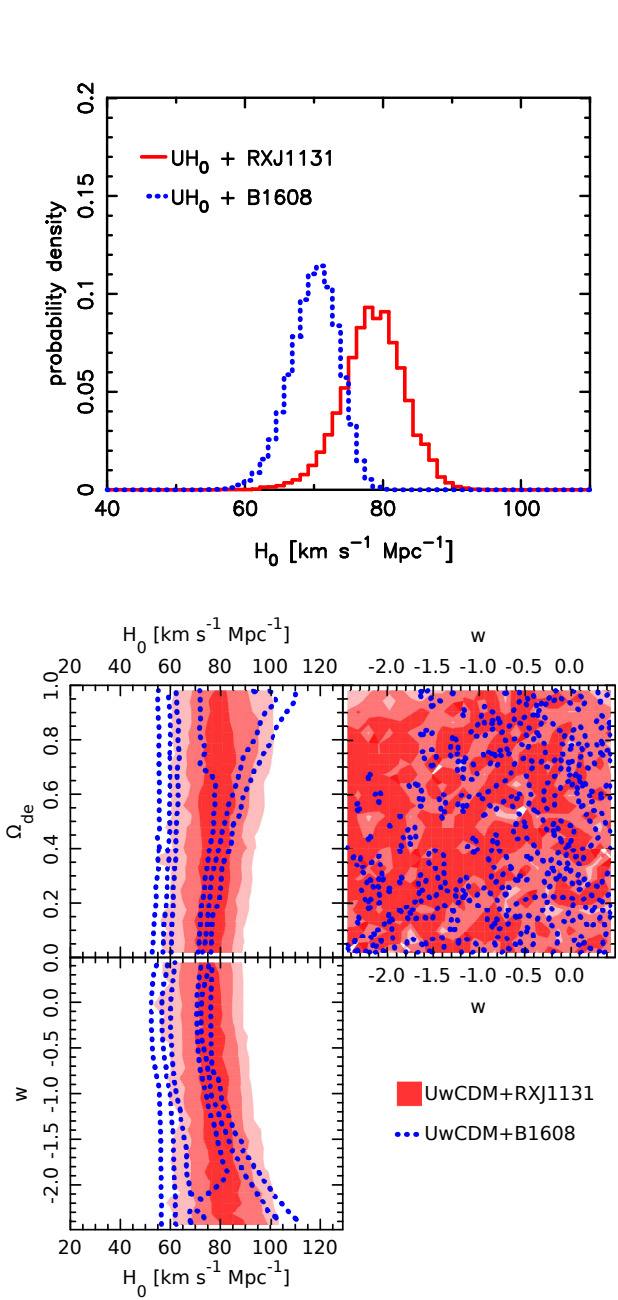}
  \caption{\label{fig:1131vs1608} Comparison of \rxj\ (solid red) with
    \blens\ (dotted blue) in U$H_0$ (top) and U$w$CDM (bottom)
    cosmologies.  The two distributions overlap within 2$\sigma$.  The
    cosmological constraints from the two lenses are statistically
    consistent with each other: the ratio of the probability that the
    two lenses share global cosmological parameters to the probability
    that the lenses require independent cosmologies is 3.2 in U$H_0$
    and 3.8 in U$w$CDM.} 
\end{figure}

How do the results of \rxj\ compare with that of \blens?  We show in
\fref{fig:1131vs1608} the overlay of the cosmological constraints of
\rxj\ and \blens\ in U$H_0$ (top panel) and U$w$CDM (bottom panel). To
investigate the consistency of the two data sets, we need to
consider their likelihood functions in the multi-dimensional
cosmological parameter space: inconsistency is defined by insufficient
overlap between the two likelihoods. We follow 
\citet{MarshallEtal06}, and compute the Bayes Factor $F$ 
in favor of a
single set of cosmological parameters and a simultaneous fit:
\be
\label{eq:hypratioL}
F = \frac{\langle L^{\rm R} L^{\rm B}\rangle}
         {\langle L^{\rm R}\rangle \langle L^{\rm B}\rangle},
\ee
where $L^{\rm R}$ and
$L^{\rm B}$ are the likelihoods of the \rxj\ and \blens\ data
respectively, computed at each prior sample point.
See the Appendix for the derivation of this result. 

For the cosmology U$H_0$, the Bayes Factor is 3.2; for U$w$CDM, it
takes the value 3.8.  For comparison, with two one-dimensional
Gaussian PDFs, $F$ takes the value of 1 when the two distributions
overlap at their 2$\sigma$ points, and is about 3.6 when they overlap
at their 1$\sigma$ point.  From this we conclude that the results from
\rxj\ and \blens\ are consistent with each other.  We do not detect
any significant residual systematics given the current uncertainties
in our measurements.

\subsection{\rxj\ and \blens\ in unison}
\label{sec:cosmo:combinelenses}

Having shown that \rxj\ and \blens\ yield consistent results with each
other, we proceed to combine the results from these two lenses for
cosmological inferences.  In particular, we consider the constraints
in the WMAP7$w$CDM and WMAP7o$\Lambda$CDM cosmologies in 
\tref{tab:cosmoprior}.  

\begin{figure}
  \centering
  \large{\bf WMAP7$w$CDM}
\\
  \includegraphics[width=0.45\textwidth, clip]{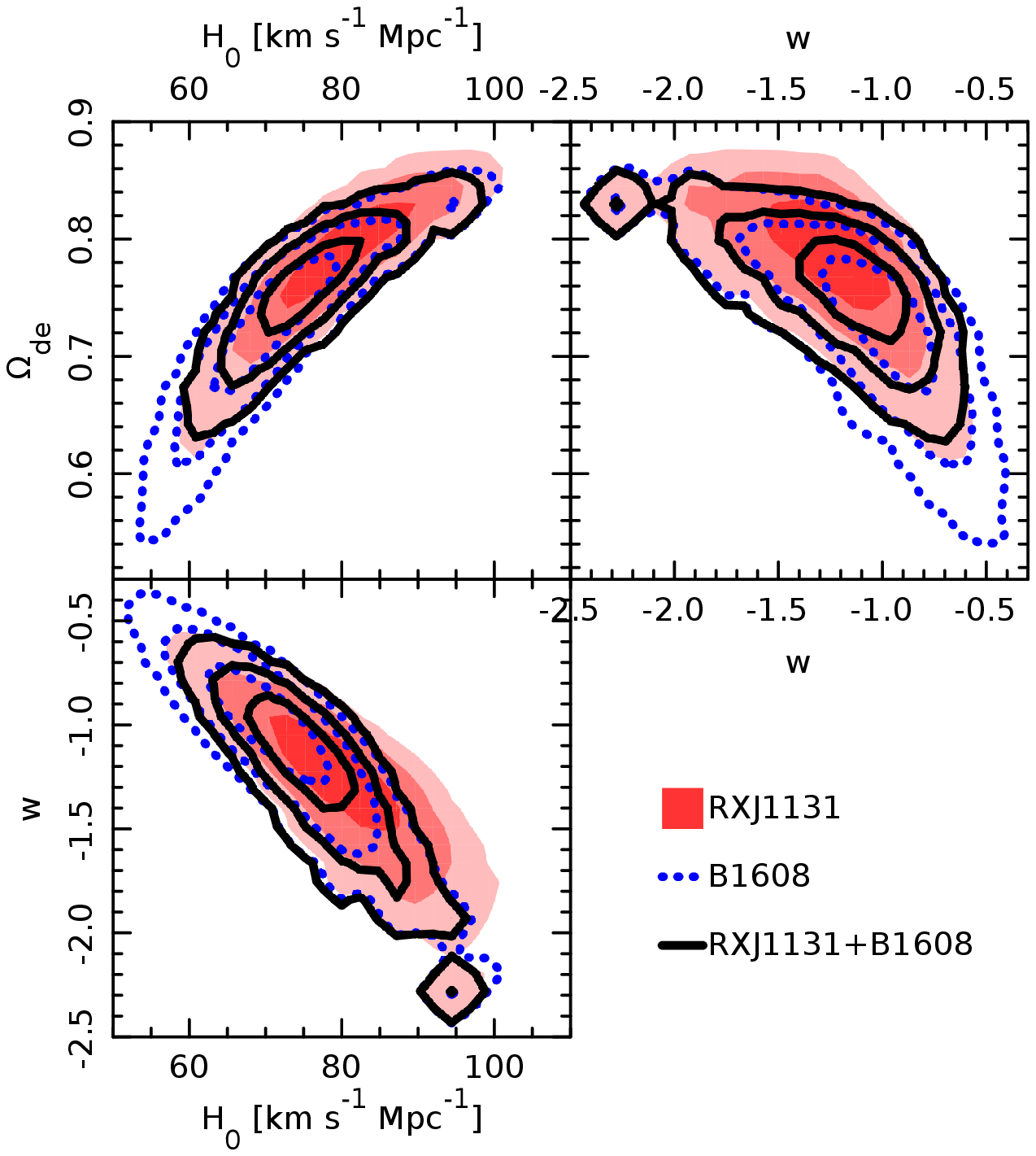}
 \\ \ \\ \large{\bf WMAP7o$\Lambda$CDM} \\
  \includegraphics[width=0.45\textwidth, clip]{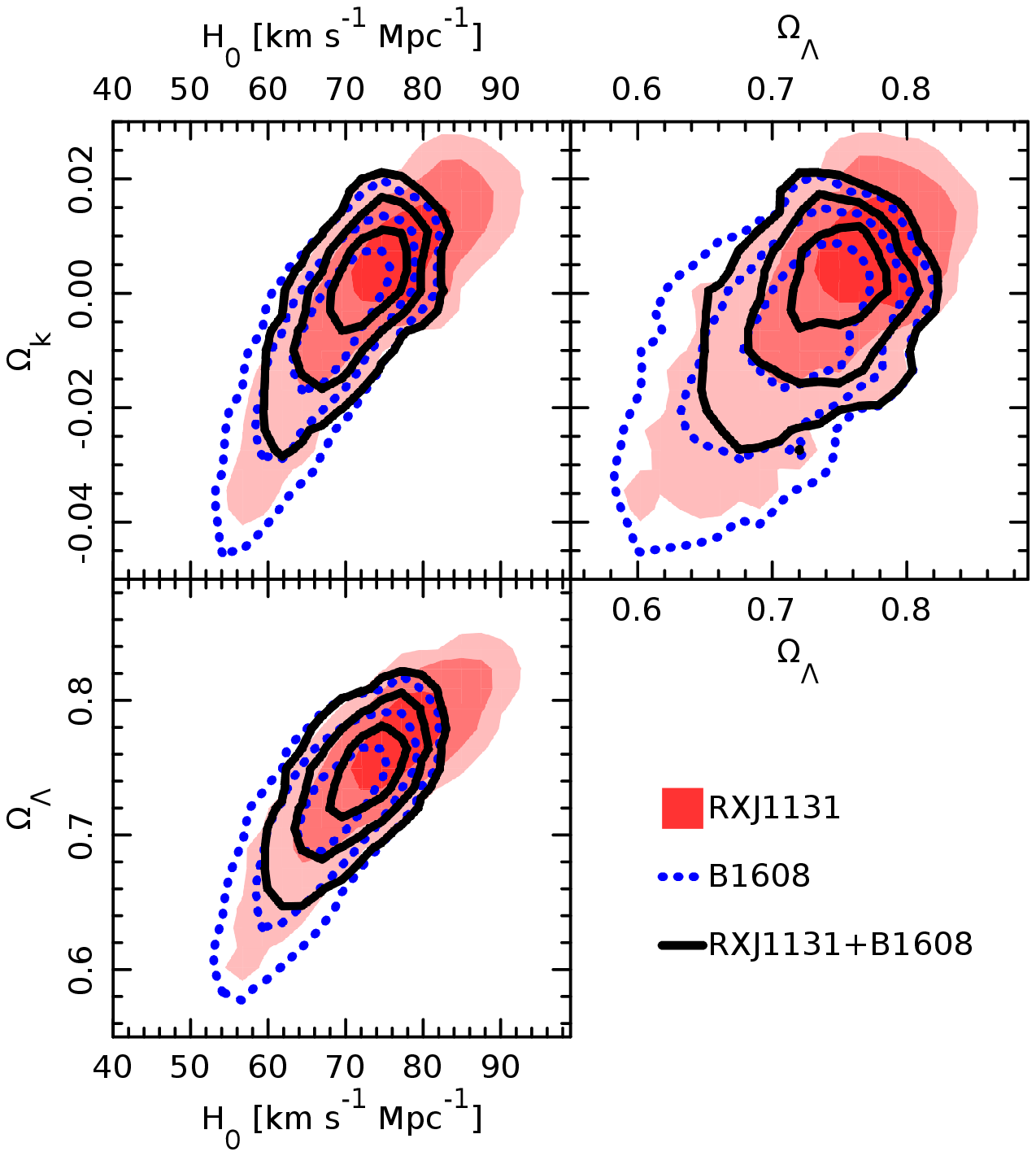}
  \caption{\label{fig:1131+1608} Cosmological constraints from the
    combination of \rxj\ and \blens\ assuming
    WMAP7$w$CDM (top) and WMAP7o$\Lambda$CDM (bottom) cosmologies.
    The combined posterior PDF is shown by the solid contours, the PDF
    for \rxj\ in 
    combination with WMAP7 is the shaded contours, and the PDF for \blens\ in
    combination with WMAP7 is the dotted contours.  Contours/shades mark
    the 68.3\%, 95.4\%, 99.7\% credible regions.}
\end{figure}

We show in \fref{fig:1131+1608} the cosmological constraints
from individual lenses in combination with WMAP7, and the combination
of both lenses and WMAP7.  By combining the two lenses, we tighten the
constraints on $H_0$, $\Ode$ and $\Ok$.  The precision on $w$ does not
improve appreciably.  With its low lens redshift, \rxj\ provide very little
information to $w$ in addition to that obtained from \blens.  In
\tref{tab:1131+1608}, we summarize the constraints from the two
lenses.  

\begin{table}
\caption{Cosmological constraints from \rxj\ and \blens\ in
  combination with WMAP7}
\label{tab:1131+1608}
\begin{center}
\begin{tabular}{lccc}
\hline
Cosmology & Parameter & Marginalized & Precision\\
 & &  Value (68\% CI) \\
\hline
& $H_0$ &  $75.2^{+4.4}_{-4.2}$ &  5.7\%\\
$w$CDM & $\Ode$ & $0.76^{+0.02}_{-0.03}$ & 2.5\%\\
& $w$ & $-1.14 ^{+0.17}_{-0.20}$ & 18\%\\
\hline
& $H_0$ &  $73.1^{+2.4}_{-3.6}$ & 4.0\%\\
o$\Lambda$CDM & $\OL$ & $0.75^{+0.01}_{-0.02}$ & 1.9\%\\
& $\Ok$ & $0.003^{+0.005}_{-0.006}$ & 0.6\% \\
\hline
\end{tabular}
\end{center}
Notes.  The $H_0$ values are in units of $\kmsMpc$.  The
``precision'' in the fourth column is defined as 
half the 68\% confidence interval, as a percentage of 75 for
$H_0$, $1$ for $\Ode$, $\OL$, and $\Ok$, and $-1.0$ for $w$.
\end{table}


\subsection{Comparison of lenses and other cosmographic probes}
\label{sec:cosmo:compprobes}

\begin{figure*}
  \centering
  \includegraphics[width=0.35\textwidth,clip, angle=270]{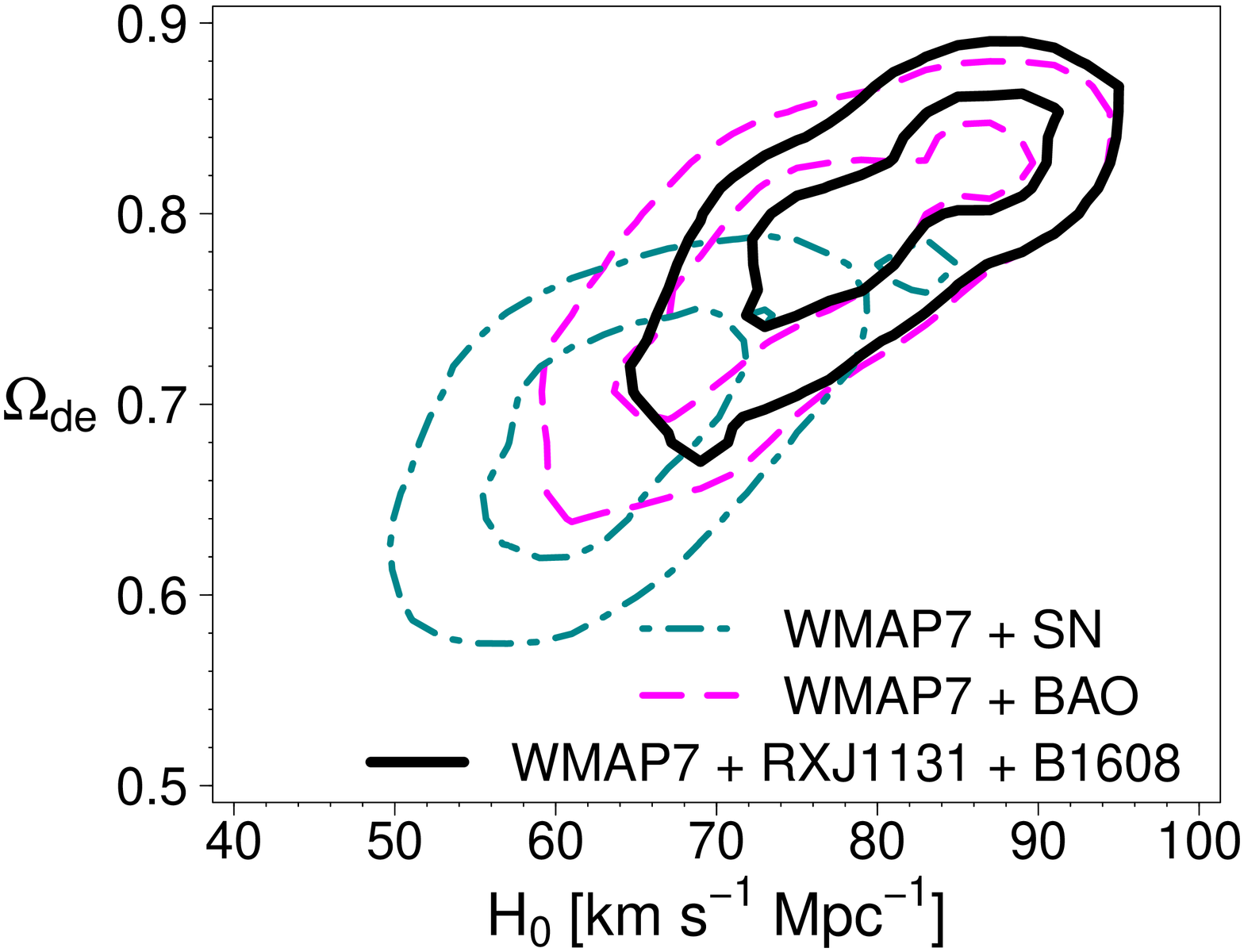} 
  \includegraphics[width=0.35\textwidth,clip, angle=270]{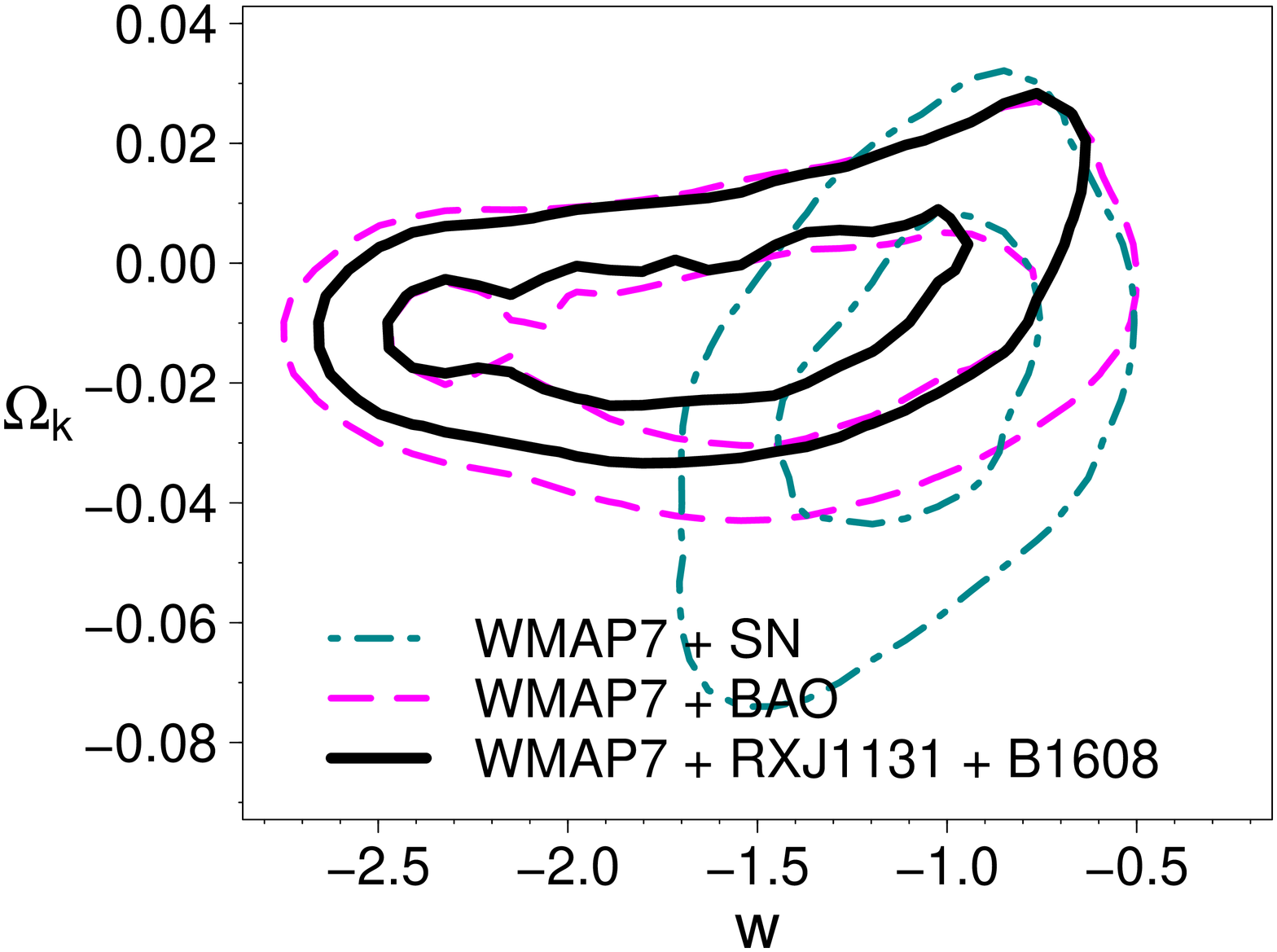} 
  \caption{\label{fig:Lenses_BAO_SN} Posterior PDF of $H_0$,
    $\Ode$, $w$ and $\Ok$ for SN (turquoise dot-dashed;
    \citet{HickenEtal09}), BAO (magenta dashed;
    \citet{PercivalEtal10}),
    time-delay lenses (black solid; this work) when each is combined with WMAP7
    in an o$w$CDM cosmology.  Contours mark the 68\%,
    and 95\% credible regions.  Time-delay lenses are highly
    complementary to other probes, particularly the CMB and SN.}
\end{figure*}

\begin{figure}
  \centering
  \includegraphics[width=0.35\textwidth, clip, angle=270]{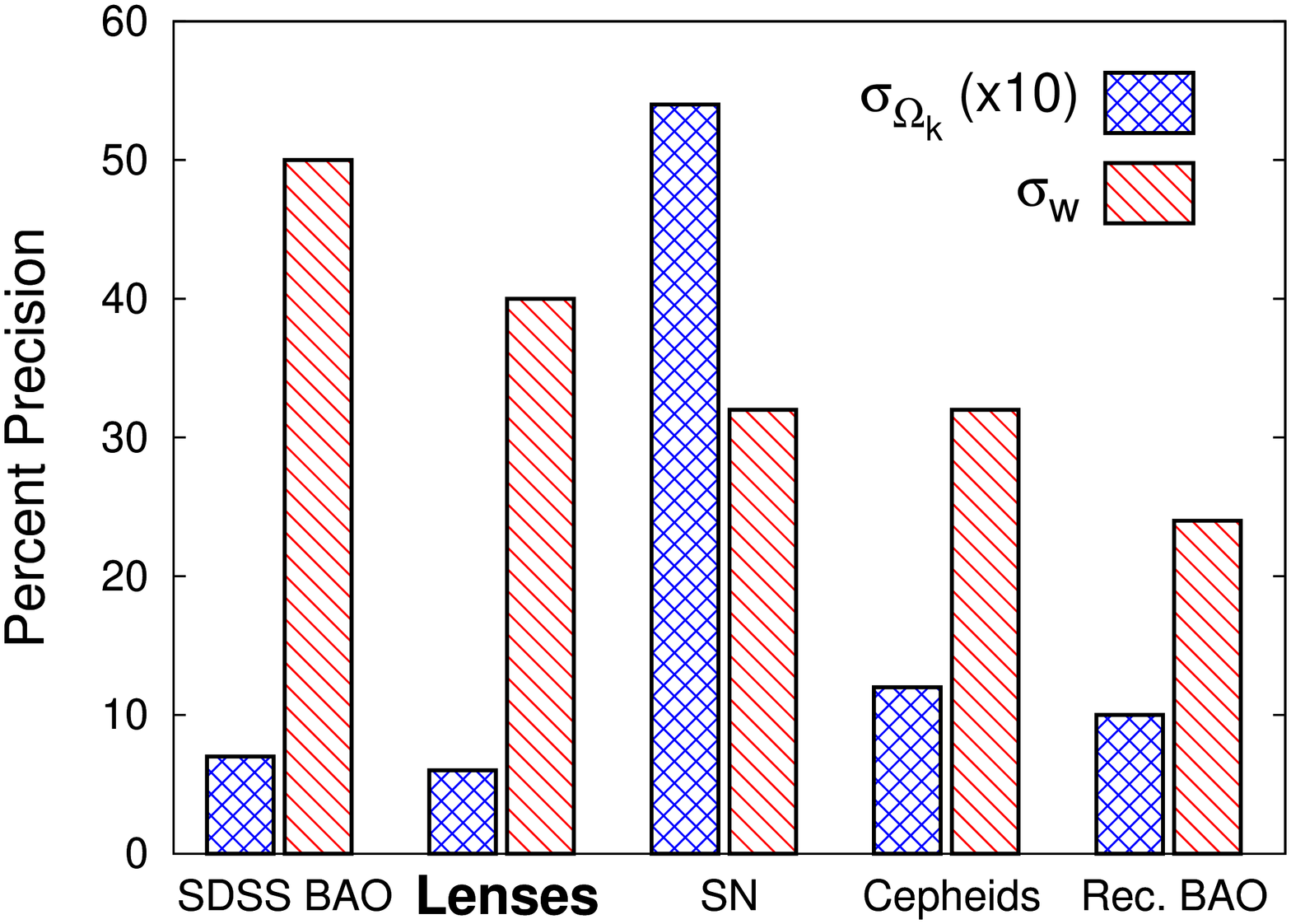}
  \caption{\label{fig:probe_precision} Precision of cosmological
    constraints on $\Ok$ and $w$ for five probes each in combination with
    WMAP in an o$w$CDM cosmology: SDSS BAO \citep{PercivalEtal10}, the
    two time-delay lenses \rxj\ and \blens\ (this work), SN
    \citep{SuzukiEtal12}, Cepheids \citep{RiessEtal11}, and
    reconstructed BAO \citep{MehtaEtal12}.  Precision for $\Ok$ and
    $w$ is defined as half the 68\% CI as a percentage of 1 and $-1$, respectively. }
\end{figure}

How do the robust time-delay distances from the strong lenses compare to
the distance measures of other probes?  We show in 
\fref{fig:Lenses_BAO_SN} a comparison of the cosmological constraints
of the two lenses, Baryon Acoustic Oscillations \citep[BAO;
e.g.,][]{PercivalEtal10, BlakeEtal11, MehtaEtal12}, and supernovae
\citep[SN; e.g.,][]{HickenEtal09, SuzukiEtal12}, when each is combined
with WMAP7 in the o$w$CDM cosmology.  The figures are qualitative since
the samples for WMAP7 chain in the o$w$CDM cosmology are sparse and we
have smoothed the contours after importance sampling.  Nonetheless, we
see that the sizes of the contours are comparable, suggesting that
even a small sample of time-delay lenses is a powerful probe of
cosmology.  Both the lenses and BAO are strong in constraining the
curvature of the Universe, while SN provides more information on the
dark energy equation of state.  Lenses are thus highly complementary
to other cosmographic probes, particularly the CMB and SN \citep[see
also, e.g.,][]{Linder11, DasLinder12}.  Each probe is consistent with
flat $\Lambda$CDM: $\Ok=0$ and $w=-1$ are within the 95\% credible
regions. 

In \fref{fig:probe_precision}, we compare the precisions on
$\Ok$ and $w$ in o$w$CDM from the following cosmological
probes in combination with WMAP: BAO
from the Sloan Digital Sky Survey (SDSS) \citep{PercivalEtal10}, 
our two time-delay lenses, SN
from the Union 2.1 sample \citep{SuzukiEtal12}, 
the Cepheids distance ladder
\citep{RiessEtal11}\footnote{To derive the constraints on $\Ok$ and
  $w$ from the combination of Cepheids and WMAP7, we importance
  sample the WMAP7 chain by a Gaussian likelihood centered on
  $H_0=73.8-1.475(w+1)\,\kmsMpc$ with a width of $2.4\,\kmsMpc$
  \citep{RiessEtal11}.  The $-1.475(w+1)$ corresponds to the tilt in
  the $H_0$-$w$ plane shown in Figure 10 of \citet{RiessEtal11}.}, and
reconstructed BAO
using the SDSS galaxies \citep{MehtaEtal12}.  We note that the precisions on the Cepheids and the
time-delay lenses are only approximates since the samples of WMAP7 are
sparse in o$w$CDM due to the large parameter space. 
Nonetheless, the histogram plot shows that time-delay lenses are a
valuable probe, especially in constraining the spatial curvature of
the Universe.


\section{Summary}
\label{sec:conclude}

We have performed a blind analysis of the time-delay lens \rxj, modeling
its high precision time delays from the COSMOGRAIL
collaboration, deep \hst\ imaging, newly measured lens
velocity dispersion, and mass contribution from line-of-sight
structures.
The data sets were combined probabilistically in a joint analysis, 
via a comprehensive model of the lens system consisting of the light of
the source AGN and its host galaxy, the light and mass of the lens
galaxies, and structures along the line of sight characterized
by external convergence and shear parameters.  
The resulting time-delay distance
measurement for the lens allows us to infer cosmological constraints.
From this study, we draw the following conclusions:
\begin{enumerate}
\item Our comprehensive lens model reproduces the global features of
  the \hst\ image and the time delays.  We quantify the uncertainty due to the deflector gravitational potential on the time-delay 
  distance to be at the $4.6\%$ level.
\item Based on the external shear strength from the lens model and the
  overdensity of galaxy count around the lens, we obtained a PDF for
  the external convergence by ray tracing through the Millennium
  Simulation.  This $\kext$ PDF contributes to the uncertainty on
  $\tdist$ also at the $4.6\%$ level.
\item Our robust time-delay distance measurement of $6\%$ takes into account
  all sources of known statistical and 
  systematic uncertainty.  We provide a fitting
  formula to describe the PDF of the time-delay distance that can be
  used to combine with any other independent cosmological probe.
\item The time-delay distance of \rxj\ is mostly sensitive to $H_0$,
  especially given the low redshift of the lens.
\item Assuming a flat $\Lambda$CDM with fixed
  $\OL=0.73$ and uniform prior on $H_0$, our 
  unblinded $H_0$ measurement from \rxj\ is $78.7^{+4.3}_{-4.5}\,\kmsMpc$.
\item The constraint on $H_0$ helps break parameter degeneracies in the CMB 
  data. In combination with WMAP7 in $w$CDM, we find
  $H_0=80.0^{+5.8}_{-5.7}\,\kmsMpc$, $\Ode=0.79\pm0.03$, and
  $w=-1.25^{+0.17}_{-0.21}$.  These are statistically consistent with
  the results from the gravitational lens \blens.  There are no
  significant residual systematics detected in our method based on
  this combined analysis of the two systems.
\item By combining \rxj, \blens\ and WMAP7, we derive the following
  constraints: $H_0=75.2^{+4.4}_{-4.2}\,\kmsMpc$,
  $\Ode=0.76^{+0.02}_{-0.03}$ and $w = -1.14^{+0.17}_{-0.20}$ in flat
  $w$CDM, and $H_0=73.1^{+2.4}_{-3.6}\,\kmsMpc$,
  $\OL=0.75^{+0.01}_{-0.02}$ and $\Ok=0.003^{+0.005}_{-0.006}$ in open
  $\Lambda$CDM.
\item Our measurement of the Hubble constant is 
  completely independent of those based on the local distance
  ladder method \citep[e.g.,][]{RiessEtal11, FreedmanEtal12},
  providing an important 
  consistency check of the standard cosmological model and of general
  relativity. 
\item A comparison of the lenses and other cosmological probes that
  are each combined with WMAP7 shows that the constraints from the
  lenses are comparable in precision to various state-of-the-art
  probes.  Lenses are particularly powerful in measuring the spatial
  curvature of the universe, and are complementary to other
  cosmological probes.
\end{enumerate}

Thanks to the dedicated monitoring by the COSMOGRAIL
\citep[e.g.,][]{VuissozEtal08, CourbinEtal11, TewesEtal12b,
  TewesEtal12a} and \citet{KochanekEtal06} collaborations,
the number of lenses with accurate and precise time delays are
increasing. 
Deep \hst\ imaging for three of these lenses will be obtained in cycle
20 to allow accurate lens mass modeling that turns the delays into
distances.  
Using the estimated uncertainties of the time-delay distances
  of the three lenses, we expect to measure $H_0$ from our assembled
  sample of five lenses (\blens, \rxj, and the three cycle 20 lenses)
  to roughly $3.8\%$ in a $w$CDM cosmology if no significant
  residual systematics are detected.
Current and upcoming telescopes and surveys including the
Panoramic Survey Telescope \& Rapid Response System, Hyper-Suprime
Camera on the Subaru Telescope, and Dark Energy Survey expect to
detect hundreds of AGN lenses with dozens of delays measured
\citep{OguriMarshall10}.  Ultimately, the Large Synoptic Survey
Telescope will discover thousands of time-delay lenses, painting a
bright future for cosmography with gravitational lens time delays.


\section*{Acknowledgments}

We thank B.~Brewer, C.~Faure, E.~Linder, and N.~Suzuki for useful
discussions.  We
are grateful to E.~Komatsu for providing us the code to compute the
likelihoods of the BAO and SN data that were used in the WMAP 7-year
analysis.  We further thank the anonymous referee whose
  detailed report and constructive comments improved the presentation
  of this work.
S.H.S.~and T.T.~gratefully acknowledge support from the Packard Foundation in the
form of a Packard Research Fellowship to T.T.. 
S.H.~and R.D.B.~acknowledge support by the National Science Foundation
(NSF) grant number AST-0807458.
P.J.M. acknowledges support from the Royal Society in the form of a research
fellowship, and is grateful to the Kavli Institute for Particle Astrophysics
and Cosmology for hosting him as a visitor during part of the period of this
investigation.
M.T., F.C., and G.M.~acknowledge support from the Swiss National Science 
Foundation (SNSF).
C.D.F.~acknowledges support from NSF-AST-0909119.
L.V.E.K.~is supported in part by an NWO-VIDI program subsidy (project 
No.\ 639.042.505).
D.S.~acknowledges support from the Deutsche Forschungsgemeinschaft,
reference SL172/1-1.
This paper is based in 
part on observations made with the NASA/ESA \textit{Hubble Space
Telescope}, obtained at the Space Telescope Science Institute, which
is operated by the Association of Universities for Research in
Astronomy, Inc., under NASA contract NAS 5-26555. These observations
are associated with program \#GO-9744.  Some of the data presented in
this paper were obtained at the W.M. Keck Observatory, which is
operated as a scientific partnership among the California Institute of
Technology, the University of California and the National Aeronautics
and Space Administration. The Observatory was made possible by the
generous financial support of the W.M. Keck Foundation.  The authors
wish to recognize and acknowledge the very significant cultural role
and reverence that the summit of Mauna Kea has always had within the
indigenous Hawaiian community.  We are most fortunate to have the
opportunity to conduct observations from this mountain.


\bibliography{ms}
\bibliographystyle{apj}


\appendix

\section{Quantifying data set consistency via the Bayes Factor $B$}
\label{sec:appendix}

\citet{MarshallEtal06} invite us to consider the following two
hypotheses: (1) $\hypG$, in which the two lenses share a common  set
of cosmological parameters $\parCosVec = \{H_0, \Ode, w\}$, and (2) $\hypI$,  in
which each of the two lenses is provided with its own independent set
of cosmological parameters,  
$\parCosVec^{\rm R} = \{H_0^{\rm R}, \Ode^{\rm R}, w^{\rm R}\}$ and
$\parCosVec^{\rm B} = \{H_0^{\rm B}, \Ode^{\rm B}, w^{\rm B}\}$, 
with which to fit the
data. Each set of parameters covers the same prior
volume as in $\hypG$.
If the two data sets are highly inconsistent,
only $\hypI$ will provide a good fit to both data sets in a joint
analysis. The question is, do the data require $\hypI$, or is
$\hypG$ sufficient?

We quantify the answer to this question with the evidence ratio, or
Bayes Factor, in favor of $\hypG$:
\be
\label{eq:hypratio}
F = \frac{P(\dR,\dB|\hypG)}{P(\dR|\hypI)\,P(\dB|\hypI)}.
\ee
where 
we have collectively
denoted all the data sets of \rxj\ as $\dR$ and of \blens\ as $\dB$.
Each of the terms on the right-hand side of the above equation can be
written in terms of a multi-dimensional
integral over the cosmological parameters.  
For example,
we have (starting with the simpler terms in the denominators)
\be
\label{eq:PdR}
P(\dR|\hypI)  =  \int {\rm d}^3\parCosVec^{\rm R}\, P(\dR| \parCosVec^{\rm R}, \hypI) \,
P(\parCosVec^{\rm R} | \hypI),
\ee
where $P(\dR | \parCosVec^{\rm R}, \hypI)$ is the likelihood of the \rxj\
data sets (the weights for the cosmological samples) that we denote by
$L^{\rm R}$.  \eref{eq:PdR} is then just the ensemble average of
the samples' likelihood values,
\be
\label{eq:dRevid}
P(\dR | \hypI) = \langle L^{\rm R} \rangle.
\ee
For $P(\dB|\hypI)$, we can rewrite the likelihood $P(\dB | \parCosVec^{\rm
  B}, \hypI)$ in terms of $\tdist^{\rm B}$ to make use of 
$P(\tdist^{\rm B}|\dB,\hypI) $ given by Equation~(35) of
\citet{SuyuEtal10}:
\be
P(\dB | \parCosVec^{\rm B}, \hypI)  = \frac{P(\tdist^{\rm
    B}(\parCosVec^{\rm B})|\dB,\hypI)\,P(\dB|\hypI)}{P(\tdist^{\rm
    B}|\hypI)}.
\ee
The ratio $Z^{\rm B}=P(\dB|\hypI)/P(\tdist^{\rm B}|\hypI)$ 
is a constant factor since the prior on $\tdist^{\rm B}$ is uniform; 
thus, we obtain
\be
\label{eq:dBevid}
P(\dB|\hypI) = Z^{\rm B} \langle L^{\rm B} \rangle,
\ee
where $L^{\rm B}$ is given by the likelihood of the time-delay distance
$P(\tdist^{\rm B}(\parCosVec^{\rm B})|\dB,\hypI)$.  

Finally, for the numerator in \eref{eq:hypratio}, we have
\bea
\label{eq:dRdBevid}
\nonumber P(\dR,\dB|\hypG) &=& \int {\rm d}^3\parCosVec \, P(\dR | \parCosVec, \hypG) \, P(\dB | \parCosVec, \hypG)\, P(\parCosVec|\hypG) \\
 & = & Z^{\rm B} \langle L^{\rm R} L^{\rm B} \rangle,
\eea
where the constant $Z^{\rm B}$ is the same as that in $P(\dB|\hypI)$
since the parameterization of the cosmology for each independent lens
is identical to that of the global cosmology (\ie\ $\parCosVec^{B}$ and
$\parCosVec$ are the same cosmological parameterization).  Substituting
Equations (\ref{eq:dRevid}), (\ref{eq:dBevid}), and
(\ref{eq:dRdBevid}) into \eref{eq:hypratio}, we obtain
\be
\label{eq:hypratioLapp}
F = \frac{\langle L^{\rm R} L^{\rm B}\rangle}
         {\langle L^{\rm R}\rangle \langle L^{\rm B}\rangle},
\ee
which can be readily computed given the values of $L^{\rm R}$ and
$L^{\rm B}$ (the weights) that we have for each cosmological sample.

\label{lastpage}
\end{document}